\documentclass[journal]{IEEEtran}

\ifCLASSINFOpdf
  \usepackage[pdftex]{graphicx}
\else
\fi

\usepackage{amsmath}
\usepackage{amsfonts}

\usepackage{multirow}
\usepackage{hyperref}

\ifCLASSOPTIONcompsoc
 \usepackage[caption=false,font=normalsize,labelfont=sf,textfont=sf]{subfig}
\else
 \usepackage[caption=false,font=footnotesize]{subfig}
\fi

\usepackage{color}

\begin{document}

\title{A Unified Speaker Adaptation Method for Speech Synthesis using Transcribed and Untranscribed Speech with Backpropagation}

\author{Hieu-Thi Luong,~\IEEEmembership{Student Member,~IEEE,}
        Junichi Yamagishi,~\IEEEmembership{Senior Member,~IEEE}
\thanks{H.-T. Luong is with the National Institute of Informatics, and with Department of Informatics, SOKENDAI (The Graduate University for Advanced Studies), Tokyo 101-8340, Japan (e-mail: luonghieuthi@nii.ac.jp)}
\thanks{J. Yamagishi is with the National Institute of Informatics, with Department of Informatics, SOKENDAI (The Graduate University for Advanced Studies), Tokyo 101-8340, Japan, and also with the Centre of Speech Technology Research, University of Edinburgh, Edinburgh EH8 9AB, U.K. (e-mail: jyamagis@nii.ac.jp).}
}

\markboth{Manuscript 2019}%
{Luong \MakeLowercase{\textit{et al.}}: A Unified Speaker Adaptation Method for Speech Synthesis}

\maketitle

\begin{abstract}
By representing speaker characteristic as a single fixed-length vector extracted solely from speech, we can train a neural multi-speaker speech synthesis model by conditioning the model on those vectors.
This model can also be adapted to unseen speakers regardless of whether the transcript of adaptation data is available or not.
However, this setup restricts the speaker component to just a single bias vector, which in turn limits the performance of adaptation process.
In this study, we propose a novel speech synthesis model, which can be adapted to unseen speakers by fine-tuning part of or all of the network using either transcribed or untranscribed speech.
Our methodology essentially consists of two steps: first, we split the conventional acoustic model into a speaker-independent (SI) linguistic encoder and a speaker-adaptive (SA) acoustic decoder; second, we train an auxiliary acoustic encoder that can be used as a substitute for the linguistic encoder whenever linguistic features are unobtainable.
The results of objective and subjective evaluations show that adaptation using either transcribed or untranscribed speech with our methodology achieved a reasonable level of performance with an extremely limited amount of data and greatly improved performance with more data.
Surprisingly, adaptation with untranscribed speech surpassed the transcribed counterpart in the subjective test, which reveals the limitations of the conventional acoustic model and hints at potential directions for improvements.
\end{abstract}


\IEEEpeerreviewmaketitle

\section{Introduction}
\IEEEPARstart{t}{hanks} to recent advances in sample-by-sample waveform generation methods \cite{van2016wavenet,mehri2016samplernn} and end-to-end models \cite{wang2017tacotron,sotelo2017char2wav}, text-to-speech (TTS) has achieved outstanding performance, with the generated speech being indistinguishable from a recording under certain conditions \cite{shen2017natural}. Due to this development, many speech-synthesis researchers have moved on to more challenging tasks, speaker adaptation being one such \cite{taigman2017voiceloop,chen2018sample}.
Speaker adaptation for speech synthesis is the task of creating a new voice for a TTS system by adjusting parameters of an initial model. Speaker adaptation is not a new topic but a well-researched one, especially for HMM-based acoustic models of speech synthesis \cite{yamagishi2009analysis} and speech recognition \cite{gales1998maximum}.
Maximum likelihood linear regression (MLLR) \cite{leggetter1995maximum,tamura1998speaker} and constrained MLLR \cite{digalakis1995speaker} are popular adaptation techniques for HMM-based systems that apply some form of linear transformation to the Gaussian distributions of the initial model.
As pointed out in \cite{yamagishi2009analysis}, there are many factors that affect performance of the adapted models besides the type of speaker transformation, such as the state of the initial model and estimation criteria.

For neural speech synthesis, training a speaker-adaptive model by conditioning on a low-dimensional speaker vector is a popular method in both multi-speaker modelling \cite{zhao2016speaker} and speaker adaptation \cite{wu2015study}.
The Deep Voice 3 model \cite{ping2017deep} adds a speaker embedding to multiple parts of the network in order to train a multi-speaker TTS model for thousands of speakers.
Arik et al. \cite{arik2018neural} used a jointly trained speaker encoder network to extract a speaker embedding of unseen speakers, while Jia et al. \cite{jia2018transfer} used a separate speaker verification network.
The Voiceloop model \cite{taigman2017voiceloop} jointly trains a speaker embedding with the acoustic model and can adapt to unseen speakers by using both the speech and transcriptions of the target speakers. Nachmani et al. \cite{nachmani2018fitting} replaced the jointly trained speaker embedding of Voiceloop with a speaker embedding obtained solely from acoustic features so that the model could adapt using untranscribed speech.
There are many reasons for performing speaker adaptation instead of conventional training, for instance reducing the speaker footprint \cite{xue2014singular} and quickly adapting to new speakers \cite{arik2018neural}. But the most important reason is its potential to handle unrefined adaptation data, whether is in an insufficient quantity \cite{chen2018sample} or unreliable quality like noisy speech \cite{takaki2018unsupervised}, incorrect transcript, or no transcript at all \cite{nachmani2018fitting}.

Even though many of the existing adaptation methods produce high quality synthetic speech with a very small amount of adaptation data \cite{chen2018sample,ping2017deep}, the similarity of their speech to the target speaker is still less than that of a well-trained single-speaker model \cite{deng2018modeling,jia2018transfer}. Moreover, the performance when adapting with only a little data is not much different than when adapting with a lot of data \cite{arik2018neural,jia2018transfer}. This is a big disadvantage especially when adapting with untranscribed speech, which is cheaper and easier to collect. To overcome these limitations, we propose a multimodal speech synthesis architecture that can adapt to unseen speakers by using transcribed or untranscribed speech.
In either case, backpropagation is used to fine tune part or all of the network. Simultaneously we investigate multiple strategies to model speaker transformations%
\footnote{We presented a proof-of-concept for using a multimodal architecture to perform speaker adaptation with untranscribed speech in \cite{luong2018multimodal}. A preliminary study on utilizing scaling and bias codes for adaption using transcribed speech is published in \cite{luong2018scaling}. The current paper combines and extends the methodologies of these two papers into a comprehensive study on speaker adaptation for speech synthesis.}. The rest of the paper is organized as follows: Section \ref{sec:related} systematically reviews the related work on speaker adaptation for neural acoustic models. Section \ref{sec:sdcomponents} describes our factorized strategies for modeling the speaker transformation, while Section \ref{sec:multimodal} explains the methodology of training and using the multimodal architecture to perform adaptation with transcribed or untranscribed speech. Section \ref{sec:expconditions} gives details about the experiments. Section \ref{sec:evaluations} shows the results of objective and subjective evaluations, and Section \ref{sec:conclusion} concludes with our finding.

\section{Related work on adapting neural acoustic models to unseen speakers}
\label{sec:related}
Speaker adaptation involves tuning the initial acoustic model using the data of unseen speakers. In the case of speech recognition, speaker adaptation makes the model perform better on the data of unseen speakers while in the case of speech synthesis, speaker adaptation allows a model to synthesize voices of new speakers.
A deep neural network (DNN) is a multilayer perceptron with many non-linear hidden layers stacked on top of each other \cite{bengio2009learning}. For speech synthesis, a typical neural acoustic model is trained to map a text representation (e.g., linguistic features) to a speech representation (e.g., acoustic features); this mapping is reversed for speech recognition. A simple feedforward hidden layer can be defined as follows:
\begin{equation}
\label{eq:fflayer}
    \boldsymbol{h}_{l} = f( \boldsymbol{W}_{l}\boldsymbol{h}_{l-1} + \boldsymbol{c}_l )
\end{equation}
where $\boldsymbol{h}_{l}$ is the output of the $l$-th hidden layer. Assuming all hidden layers have the same $m$ hidden units, the parameters of the $l$-th hidden layer are a weight matrix $\boldsymbol{W}_l \in \mathbb{R}^{m \times m}$ and a bias vector $\boldsymbol{c}_l \in \mathbb{R}^{m \times 1}$. $f(.)$ is an element-wise activation function with non-linear functions being the most common type.
The speaker-dependent layer or speaker layer is one whose parameters have been trained on data of one specific speaker:
\begin{equation}
\label{eq:sdlayer}
    \boldsymbol{\hat{h}}_{l} = f( \boldsymbol{W}^{(k)}_{l}\boldsymbol{h}_{l-1} + \boldsymbol{c}^{(k)}_l )
\end{equation}
where $\boldsymbol{\hat{h}}_{l}$ represents the speaker layer with parameters $\boldsymbol{W}^{(k)}_{l}$ and $\boldsymbol{c}^{(k)}_l$ depending on $k$-th speaker. The conventional single speaker speech synthesis model is essentially a neural network with all of its layers trained on data of a single speaker.

Training or fine-tuning the entire neural network is a simple and straightforward approach to obtain a speech synthesis model for a target speaker.
However, it is vulnerable to overfitting when the target speaker has a limited amount of data, as there are too many parameters to adjust.
Many adaptation techniques have been proposed to overcome this problem. Below, we systematically review them by characterizing them according to three factors: 1) the components used to model the speaker characteristics; 2) the speaker awareness (or unawareness) of the initial model and 3) the ability to perform adaptation using untranscribed speech.

\subsection{Speaker component}
\label{subsec:speakercomponents}

The speaker component is the most crucial aspect of a speaker adaptation methodology as it directly affects the speaker footprint and performance of the adapted model.
Here, we could adapt either the entire neural network \cite{kons2018neural} or all but the output layer \cite{li2018speaker} of a pre-trained model.
However, as mentioned above, this approach is vulnerable to overfitting, so techniques like regularization \cite{yu2013kl} or early stopping \cite{chen2018sample} are often introduced in the adaptation stage. 
Instead of using regularization, the number of adaptable parameters could be reduced as a way to prevent overfitting. The speaker component can be reduced to just one \cite{fan2015multi} or a few layers \cite{huang2018linear}. These layers can be further factorized \cite{zhao2016low} to discourage the adapted model of the target speaker from straying too far from the initial state. Below, we categorize these factorized methods on the basic of the type of transformation they model within a single token layer of the neural network:

\subsubsection{Speaker layer}
We can use the entire layers with both weights and biases as the speaker components. The Equation \ref{eq:sdlayer} described a simple speaker layer approach.
Usually, these speaker layers are strategically placed at input \cite{neto1995speaker}, output \cite{li2016multi} or in-between the hidden layers \cite{huang2018linear} depending on the task at hand.
The weights and biases can be factorized in various ways to further reduce the speaker footprint \cite{zhao2016low,huang2018linear}.

\subsubsection{Speaker weight}
Many approaches only use the layer weights as the speaker components. For example singular value decomposition (SVD) bottleneck \cite{xue2014singular} factorizes the full matrix $\boldsymbol{W}_l^{(k)} \in \mathbb{R}^{m \times m}$ into products of several low-rank matrices to reduce the speaker footprint:
\begin{equation}
\label{eq:svdlayer}
    \boldsymbol{\hat{h}}_{l} = f( \boldsymbol{U}_{l}\boldsymbol{A}_{l}^{(k)}\boldsymbol{V}_{l}\boldsymbol{h}_{l-1} + \boldsymbol{c}_l )
\end{equation}
where $\boldsymbol{U}_{l} \in \mathbb{R}^{m \times n}$ , $\boldsymbol{V}_{l} \in \mathbb{R}^{n \times m}$ and $\boldsymbol{A}_{l}^{(k)} \in \mathbb{R}^{n \times n}$. By setting $n \ll m$\, the speaker specific parameters become a lot less numerous than using the square matrix $\boldsymbol{W}_l^{(k)}$.
Similarly, in the cluster adaptive training (CAT) method proposed by Tan et al. \cite{tan2015cluster}, the speaker weight is estimated based on an interpolation between several canonical matrices and hence the interpolation coefficients $\boldsymbol{\lambda}_l^{(k)} \in \mathbb{R}^p$ are speaker-specific parameters:
\begin{equation}
\label{eq:cat}
 \boldsymbol{W}_l^{(k)} = \sum_{i=1}^p\lambda_{l,i}^{(k)}\boldsymbol{W}_{l,i}
\end{equation}
where $\boldsymbol{W}_l^{(k)}$ depends on the canonical set $\boldsymbol{M}_l = \{\boldsymbol{W}_{l,1},..., \boldsymbol{W}_{l,p}\}$. Factorized hidden layer (FHL) \cite{samarakoon2016factorized} exercises a similar concept, modeling the speaker weight as a subspace over a finite set of canonical matrices.

\subsubsection{Speaker scaling}
Scaling is the most common type of linear transformation used to model speaker transformation by itself. For example,
Learning hidden unit contribution (LHUC) \cite{swietojanski2014learning} uses a speaker-dependent vector $\boldsymbol{a}^{(k)}_l \in \mathbb{R}^{m \times 1}$ to adjust the output of the hidden layers:
\begin{equation}
\label{eq:lhuc}
    \boldsymbol{\hat{h}_l} = \boldsymbol{a}^{(k)}_l \circ f( \boldsymbol{W}_{l}\boldsymbol{h}_{l-1} + \boldsymbol{c}_l ).
\end{equation}
From the perspective of the next hidden layer, $\boldsymbol{a}^{(k)}_l$ is basically a diagonal scaling matrix: 
\begin{equation}
\label{eq:lhuc_rewrite}
    \boldsymbol{\hat{h}}_{l+1} = f( \boldsymbol{W}_{l+1} {\rm diag} (\boldsymbol{a}^{(k)}_l) \boldsymbol{h}_{l} + \boldsymbol{c}_{l+1} )
\end{equation}
where ${\rm diag}$ is the operation of changing an ${m \times 1}$ vector into a diagonal ${m \times m}$ matrix. By restricting the speaker transformation to just scaling, it reduces the speaker footprint as well as prevents the adapted model from deviating too far from the initial state.
Just like the speaker weight, speaker scaling can be factorized further using the subspace approach.
Samarakoon et al. \cite{samarakoon2016subspace} proposed subspace LHUC, in which $\boldsymbol{a}^{(k)}_l$ is projected from a vector $\boldsymbol{s}^{a,(k)}_l \in \mathbb{R}^{p \times 1}$ of arbitrary size $p$ by using a SI matrix $\boldsymbol{W}^a_l \in \mathbb{R}^{m \times p}$:   
\begin{equation}
    \boldsymbol{a}^{(k)}_l = 2 \times \sigma( \boldsymbol{W}^a_l \boldsymbol{s}^{a,(k)}_l)
\end{equation}
Other variations of subspace speaker scaling are investigated in \cite{luong2018scaling} and \cite{cui2017embedding}.

\subsubsection{Speaker bias}
The layer bias has also been proven to be an effective speaker-specific parameter to be used on its own \cite{zhao2016speaker,luong2017adapting}.
In practice, to model a speaker bias, we augment the input or hidden layer(s) with a one-hot vector representing the speaker \cite{hojo2018dnn}:
\begin{equation}
\label{eq:speakerbias}
    \boldsymbol{\hat{h}}_{l} = f( \boldsymbol{W}_{l}\boldsymbol{h}_{l-1} + \boldsymbol{c}_l + \boldsymbol{b}_l^{(k)} )
\end{equation}
where $\boldsymbol{b}_l^{(k)} \in \mathbb{R}^{m \times 1}$ is a speaker-specific bias projected from the speaker one-hot vector. We could factorize the speaker bias further by using a continuous vector to represent speaker instead of using the discrete one-hot vector. Abdel-Hamid et al.\ \cite{abdel2013fast} jointly train the speaker embedding with the acoustic model, whereas Saon et al. \cite{saon2013speaker} use i-vector obtained from an external system to represent speaker:
\begin{equation}
\label{eq:biascodes}
    \boldsymbol{b}_l^{(k)} = \boldsymbol{W}^{b}_l\boldsymbol{s}_l^{b,(k)}
\end{equation}
where $\boldsymbol{W}^{b}_l \in \mathbb{R}^{m \times q}$ is a subspace matrix that projects an arbitrary-sized vector $\boldsymbol{s}^{b,(k)} \in \mathbb{R}^{q \times 1}$ to the speaker bias.

\subsubsection{Combinations}
By categorizing the adaptation methods based on the type of transformation within a single layer, we got a comprehensive overview about the nature of the speaker components. However the hierarchical nature of the neural network adds another aspect for speaker modeling, as these speaker components can be used together at one or multiple layers.
For example, as the speaker scaling and bias complement each other, certain representations of them are utilized for speaker modeling in \cite{luong2018scaling} and \cite{cui2017embedding}. Similarly, the subspace variations of speaker weight and bias are investigated in \cite{samarakoon2016factorized}. In \cite{wu2015study} and \cite{takaki2016speaker} the effects of combining several types of speaker components are investigated. It is difficult to provide an absolute answer about what the best setup for adaptation is, as the performance depends heavily on the network architecture, quality of the speech data, ratio between the speaker footprint and the amount of adaptation data, as well as the training conditions of the initial model.

\subsection{Speaker-awareness of the initial model}

Besides the speaker component, the state of the initial model also affects the performance of the adapted model \cite{yamagishi2009analysis}.
More specifically, we can classify the initial model either as speaker-aware or speaker-unaware.
A speaker-unaware model is trained without information about the speaker.
This sort of model includes the conventional SI model of speech recognition and the single-speaker \cite{zen2013statistical,kons2018neural} or average-voice model \cite{gutkin2016tts} of speech synthesis.
A speaker-aware model is trained with the speaker components integrated into the initial model.
For speech recognition, it is generally known as speaker-adaptive training (SAT) \cite{saon2013speaker}. For speech synthesis, it is the multi-speaker model \cite{zhao2016speaker}.

Most speaker components reviewed in Section \ref{subsec:speakercomponents} can be used for both speaker-aware and speaker-unaware setups. For example, Fan et al. \cite{fan2015multi} train a multi-speaker model with a speaker output layer, capable of adapting to an unseen speaker by fine-tuning a new layer for the target.
Meanwhile Huang et al. \cite{huang2018linear} add new layers for unseen speakers on top of a pretrained speaker-unaware single-speaker model.
Similar to the LHUC method, Swietojanski et al. \cite{swietojanski2014learning} proposed to add speaker parameters on top of the SI model for adaptation. By constrast, in a more recent publication, they introduced SAT-LHUC \cite{swietojanski2016sat} which adds LHUC parameters right from the training stage.
The speaker awareness or unawareness of the initial model does not change the structure of the adapted model, but it changes the representation learned by the hidden layers. Training a speaker-aware model encourages the model to disentangle speaker characteristics (style) from the linguistic information (content) which in turn would help the adaptation \cite{swietojanski2016sat}.

\subsection{Adaptation using untranscribed speech}

As a neural network is trained with the backpropagation algorithm \cite{rumerlhar1986learning}, we can use backpropagation to fine-tune part of or all of the model in the adaptation stage as well if both input features and output features of the target domain are available.
Therefore, it is straightforward to adapt the acoustic model to an unseen speaker when the adaptation data is transcribed speech, whereas it becomes trickier when we have untranscribed speech.
In automatic speech recognition (ASR), adaptation using speech and text is referred to as supervised adaptation while using only speech is referred to as unsupervised adaptation \cite{yu2013kl,liao2013speaker}; we will adopt this terminology for speech synthesis in the rest of this paper.
The common unsupervised adaptation approach for ASR is the two-pass adaptation: an SI ASR model is used to obtain the text label; then speech and the prediction label are used to perform unsupervised adaptation with backpropagation just like in the supervised counterpart \cite{yu2013kl,li2018speaker}.
As both supervised and unsupervised adaptation are based on backpropagation, we could use any type of speaker component reviewed in Section \ref{subsec:speakercomponents}.

In the case of speech synthesis, a common method of unsupervised speaker adaptation is to assume that the characteristics of the $k$-th speaker can be represented by a single fixed-length vector $\boldsymbol{s}^{b,(k)} \in \mathbb{R}^{q \times 1}$ extracted solely from speech. The speaker-adaptive model is then trained by using $\boldsymbol{s}^{b,(k)}$ as a bias code:
\begin{equation}
\label{eq:spkembedding}
    \boldsymbol{\hat{h}}_{l} = f( \boldsymbol{W}_{l}\boldsymbol{h}_{l-1} + \boldsymbol{c}_l + \boldsymbol{W}^{b}_l\boldsymbol{s}^{b,(k)})
\end{equation}
To perform adaptation for unseen speakers, we simply extract $\boldsymbol{s}^{b,(k)}$ from the speech of the target by using an external system. This approach is sometimes referred to as one-shot learning \cite{tjandra2018machine}, as it does not involve an optimization loop with backpropagation.
For example, Wu et al. used i-vectors as the bias code \cite{wu2015study}, while Doddipatla et al. \cite{doddipatla2017speaker} and Jia et al. used d-vectors \cite{jia2018transfer}. Tjandra et al. \cite{tjandra2018machine} jointly trained a DeepSpeaker network with the acoustic model, which is used to extract the speaker vector.
By restricting the speaker component to a single bias code, we also restrict the performance of the adaptation. This leads to a gap in performance between the seen and unseen speakers \cite{luong2017adapting,jia2018transfer}.
To overcome this limitation, we proposed the multimodal speech synthesis architecture in our previous study \cite{luong2018multimodal}, which allows both supervised and unsupervised speaker adaptation to be conducted using backpropagation algorithm. In this paper, we improve upon this methodology.

\section{Speaker components for modelling speaker-adaptive speech synthesis systems}
\label{sec:sdcomponents}
\subsection{Speaker scaling and bias for speaker-adaptive modeling and adapting to unseen speakers}
\label{subsec:proposal-scaling-bias}
We conducted a preliminary study on using scaling and bias codes in \cite{luong2018scaling}. The results showed that having both components does improve the performance of speaker adaptation. However it does not seem to yield further improvement when more data become available \cite{luong2018scaling}.
To address this limitation, we can increase the amount of adaptable parameters by allowing each layer to have its own speaker scaling and bias: 
\begin{equation}
\label{eq:speakerff}
    \boldsymbol{\hat{h}}_{l} = f( {\rm diag}(\boldsymbol{a}^{(k)}_{l}) \boldsymbol{W}_{l} \boldsymbol{h}_{l-1} + \boldsymbol{c}_l + \boldsymbol{b}_l^{(k)} ) 
\end{equation}
where $\boldsymbol{a}^{(k)}_{l} \in \mathbb{R}^{m \times 1}$ and $\boldsymbol{b}^{(k)}_{l} \in \mathbb{R}^{m \times 1}$ are speaker-specific scaling and bias at the $l$-th layer. Compared with LHUC \cite{swietojanski2014learning} described by Equation \ref{eq:lhuc_rewrite}, our scaling operation is placed on the other side of the layer weight. There is no significance in this choice besides that we want both $\boldsymbol{a}^{(k)}_{l}$ and $\boldsymbol{b}^{(k)}_{l}$ to inherit the number of units of the $l$-th host layer.
Moreover, to prevent overfitting while still having speaker components in multiple layers we can apply the subspace approach to factorize the speaker scaling and bias into scaling and bias codes:
\begin{align}
\label{eq:speakercodes}
    \boldsymbol{a}^{(k)}_{l} &= \boldsymbol{W}^{a}_l\boldsymbol{s}^{a,(k)}_l \\
    \boldsymbol{b}^{(k)}_{l} &= \boldsymbol{W}^{b}_l\boldsymbol{s}^{b,(k)}_l
\end{align}
where $\boldsymbol{s}^{a,(k)}_l \in \mathbb{R}^{p \times 1}$ and $\boldsymbol{s}^{b,(k)}_l \in \mathbb{R}^{q \times 1}$ are scaling and bias codes of the $l$-th layer and have arbitrary size $p$ and $q$. These codes are projected into a speaker scaling and bias $\boldsymbol{a}^{(k)}_{l}$, $\boldsymbol{b}^{(k)}_{l}$ by using the SI matrices $\boldsymbol{W}^{a}_l \in \mathbb{R}^{m \times p}$ and $\boldsymbol{W}^{b}_l \in \mathbb{R}^{m \times q}$ trained with data of multiple speakers in the training stage. This reduces the number of parameters to be fine-tuned in adaptation stage.

Next, we extend the definitions of the speaker scaling and bias to gated convolution layers \cite{van2016conditional} which we use in our experiments to capture the temporal context in the time domain and help the information flow \cite{srivastava2015training}:
\begin{multline}
    \label{eq:speakergate}
    \boldsymbol{\hat{h}}_{l} = \tanh( {\rm diag}(\boldsymbol{a}^{f,(k)}_{l}) \boldsymbol{W}^{f}_{l} \boldsymbol{h}_{l-1} + \boldsymbol{c}^f_l +\boldsymbol{b}_l^{f,(k)} )  \\
    \odot \sigma( {\rm diag}(\boldsymbol{a}^{g,(k)}_{l}) \boldsymbol{W}^{g}_{l} \boldsymbol{h}_{l-1} + \boldsymbol{c}^g_l + \boldsymbol{b}_l^{g,(k)} )
\end{multline}
where the filter and gate have their own scaling vectors $\boldsymbol{a}^{f,(k)}_{l},\, \boldsymbol{a}^{g,(k)}_{l} \in \mathbb{R}^{m \times 1}$ and bias vectors $\boldsymbol{b}^{f,(k)}_{l},\, \boldsymbol{b}^{g,(k)}_{l} \in \mathbb{R}^{m \times 1}$. Just as in the case of the feedforward layer we can factorize the speaker vectors into smaller speaker scaling and bias codes:
\begin{align}
    \boldsymbol{a}^{f,(k)}_{l} &= \boldsymbol{W}^{a,f}_l\boldsymbol{s}^{a,(k)}_l & \text{a}&\text{nd} & \boldsymbol{a}^{g,(k)}_{l} &=  \boldsymbol{W}^{a,g}_l\boldsymbol{s}^{a,(k)}_l \\
    \boldsymbol{b}^{f,(k)}_{l} &= \boldsymbol{W}^{b,f}_l\boldsymbol{s}^{b,(k)}_l & \text{a}&\text{nd} & \boldsymbol{b}^{g,(k)}_{l} &= \boldsymbol{W}^{b,g}_l\boldsymbol{s}^{b,(k)}_l
\end{align}
where the scaling code $\boldsymbol{s}^{a,(k)}_l \in \mathbb{R}^{p \times 1}$ and bias code $\boldsymbol{s}^{b,(k)}_l \in \mathbb{R}^{q \times 1}$ are shared between the filter and the gate.

\subsection{Adapting to unseen speakers by fine-tuning entire speaker-adaptive network}
\label{subsec:method-finetune}

By adding a speaker scaling along with the speaker bias, we can model more sophisticated transformation than just using speaker bias. However, it is still restricted when comparing with using a speaker layer.
Recent studies \cite{chen2018sample,deng2018modeling} have shown that fine-tuning the entire network along with the speaker embedding is better than fine-tuning just the speaker-embedding.
Given a SA layer with the speaker bias code defined by Equation \ref{eq:spkembedding}, by finetuning the SA layer we obtained an adapted layer defined as follows:
\begin{equation}
\label{eq:adaptall}
    \boldsymbol{\hat{h}}_{l} = f( \boldsymbol{W}^{(k)}_{l}\boldsymbol{h}_{l-1} + \boldsymbol{c}^{(k)}_l + \boldsymbol{W}^{b,(k)}_l\boldsymbol{s}^{(k)})
\end{equation}
where all parameters of the layer now depend on the target speaker. However it is redundant to have $\boldsymbol{c}^{(k)}_l + \boldsymbol{W}^{b,(k)}_l\boldsymbol{s}^{(k)}$ model a speaker bias as a single vector $\boldsymbol{c}^{(k)}_l \in \mathbb{R}^{m \times 1}$ can perform a same job.

Based on the above observations, we proposed a similar adaptation strategy in which we fine-tune entire layers of an initial SA network by first removing all speaker components like $\boldsymbol{b}_l^{(k)}$ and ${\rm diag}(\boldsymbol{a}^{(k)}_{l})$.
The final adapted layer has a structure described by the Equation \ref{eq:sdlayer}. Liu et al. \cite{liu2018wavenet} used a similar strategy to adapt a multi-speaker Wavenet vocoder \cite{hayashi2017investigation} to unseen speakers with limited data.
We hypothesize that a speaker-aware model with all speaker-specific parameters stripped is a good initialization for the adaptation.

\section{Multimodal architecture for unsupervised speaker adaptation}
\label{sec:multimodal}

\begin{figure}[tb]
  \centering
  \includegraphics[width=0.85\columnwidth]{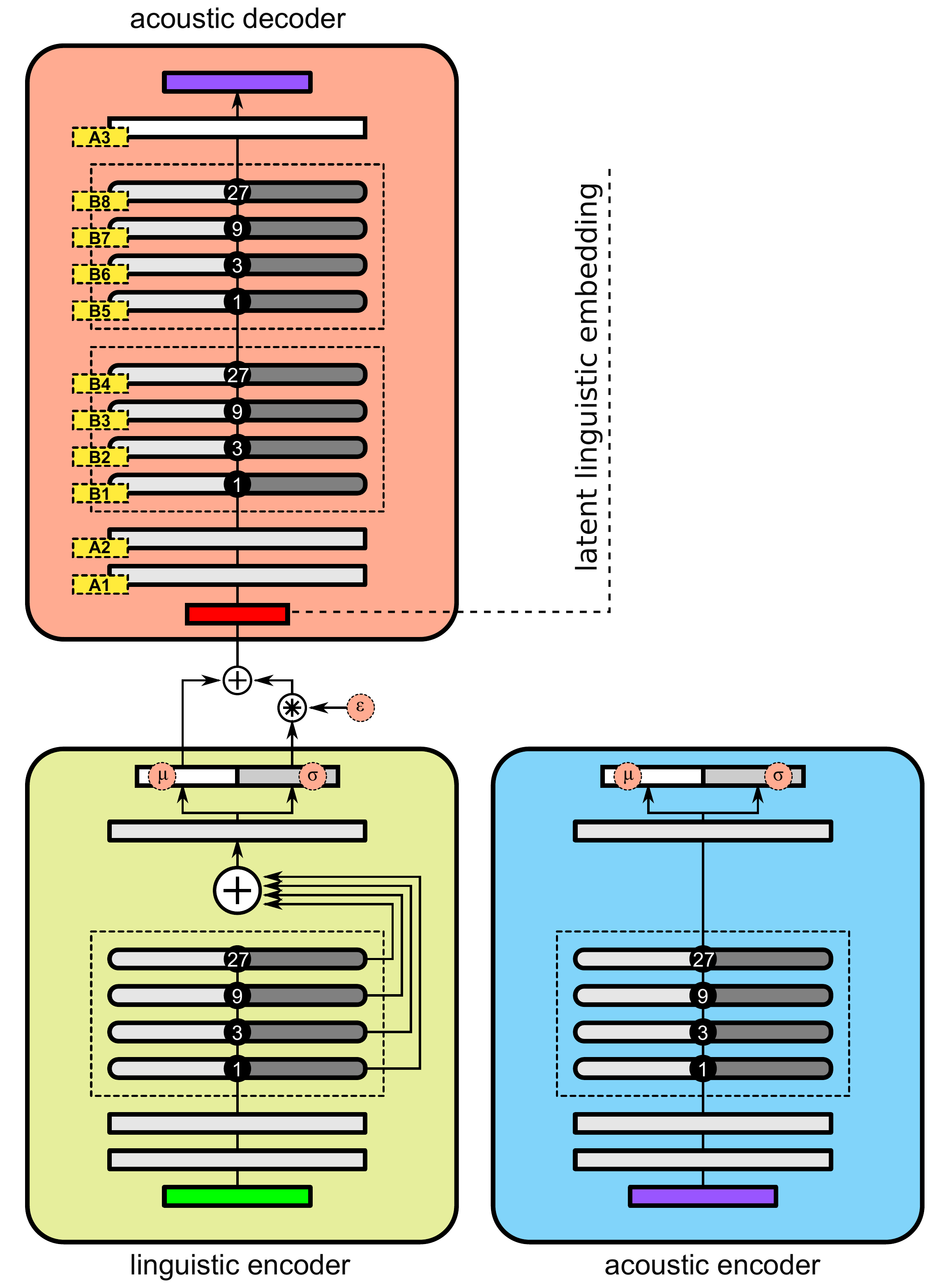}
\caption{
Blueprint of proposed multimodal architecture used in the experiments. The layers that can potentially contain speaker components are marked with a yellow identifier tag. Layers with numbers in the middle are dilated convolution layers; the number indicates the dilation rate.}
\vspace{-3mm}
\label{fig:architecture}
\end{figure}

We proposed a novel method for unsupervised speaker adaptation in our previous publication \cite{luong2018multimodal}. The main idea is splitting the conventional acoustic model into an SI linguistic encoder and SA common layers and then training an auxiliary speech encoder to be used as substitute for the linguistic encoder when linguistic feature is unobtainable, so that the adaptation can still be conducted with backpropagation.
The proposed method has several limitations: we have to use waveform as the input of the speech encoder, as the network tends to ignore speaker embedding when we use acoustic features; the quality of the generated speech is still low in general.
In this paper, we refine the methodology proposed in \cite{luong2018multimodal}. The enhanced multimodal architecture is illustrated in Figure \ref{fig:architecture}, with the three modules renamed as linguistic encoder, acoustic encoder, and acoustic decoder.
The biggest change is that the encoders no longer output a deterministic latent variable, namely a latent linguistic embedding (LLE), but a density function of it. The idea is inspired by mixture density network (MDN) \cite{bishop1994mixture,zen2014deep} and variational autoencoder (VAE) \cite{larsen2015autoencoding}. By modeling the density function of LLE, we encourage the network to learn a continuous latent space for it.

\begin{figure*}[t]
  \centering
  \captionsetup[subfigure]{justification=centering}
  \subfloat[Training \label{fig:stage-training}][Training\\(Jointly train all modules)]{\includegraphics[height=0.48\linewidth]{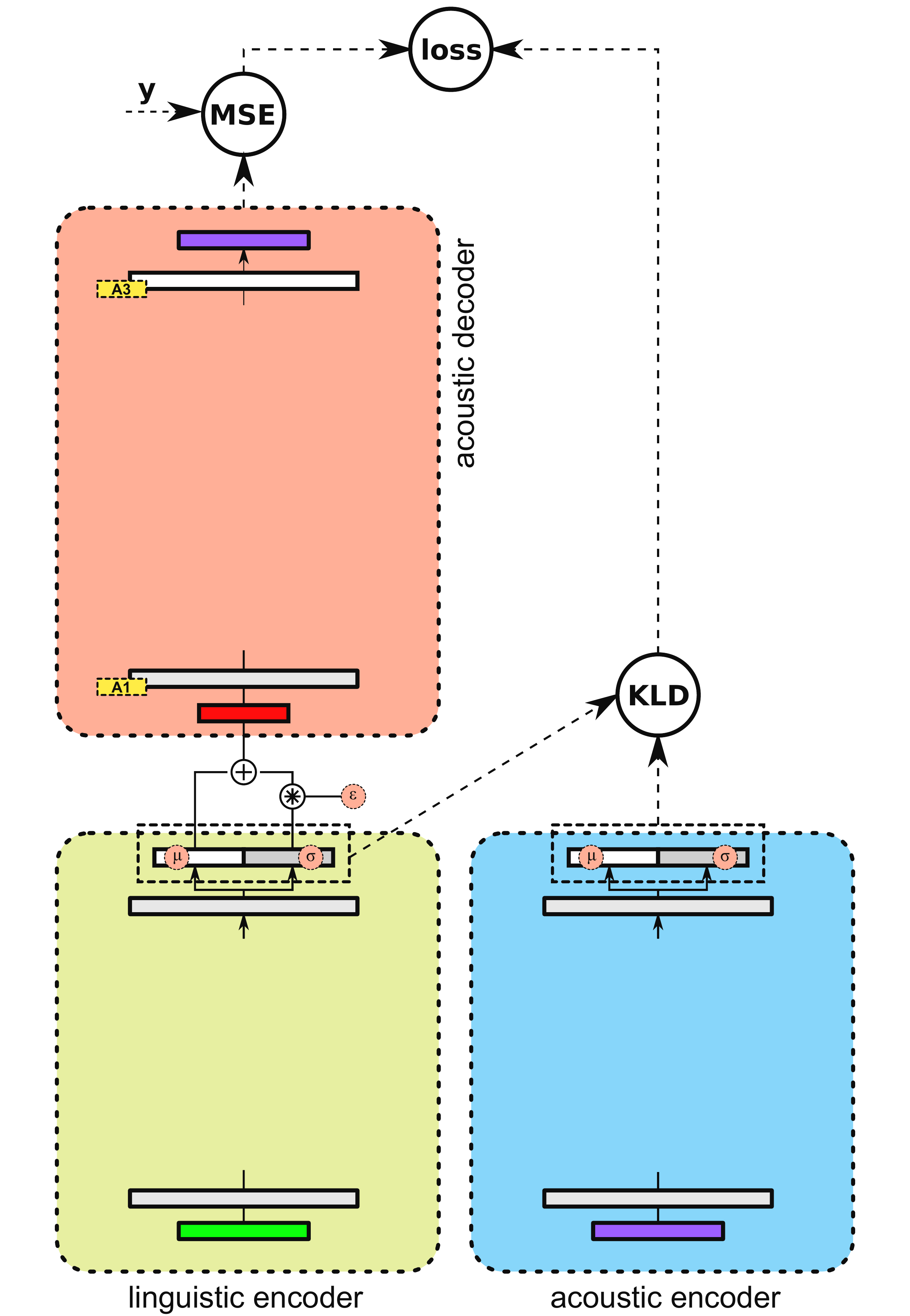}}
  \hfil
  \subfloat[Inference\label{fig:stage-inference}][Inference\\(Text-to-Speech stack)]{\includegraphics[height=0.48\linewidth]{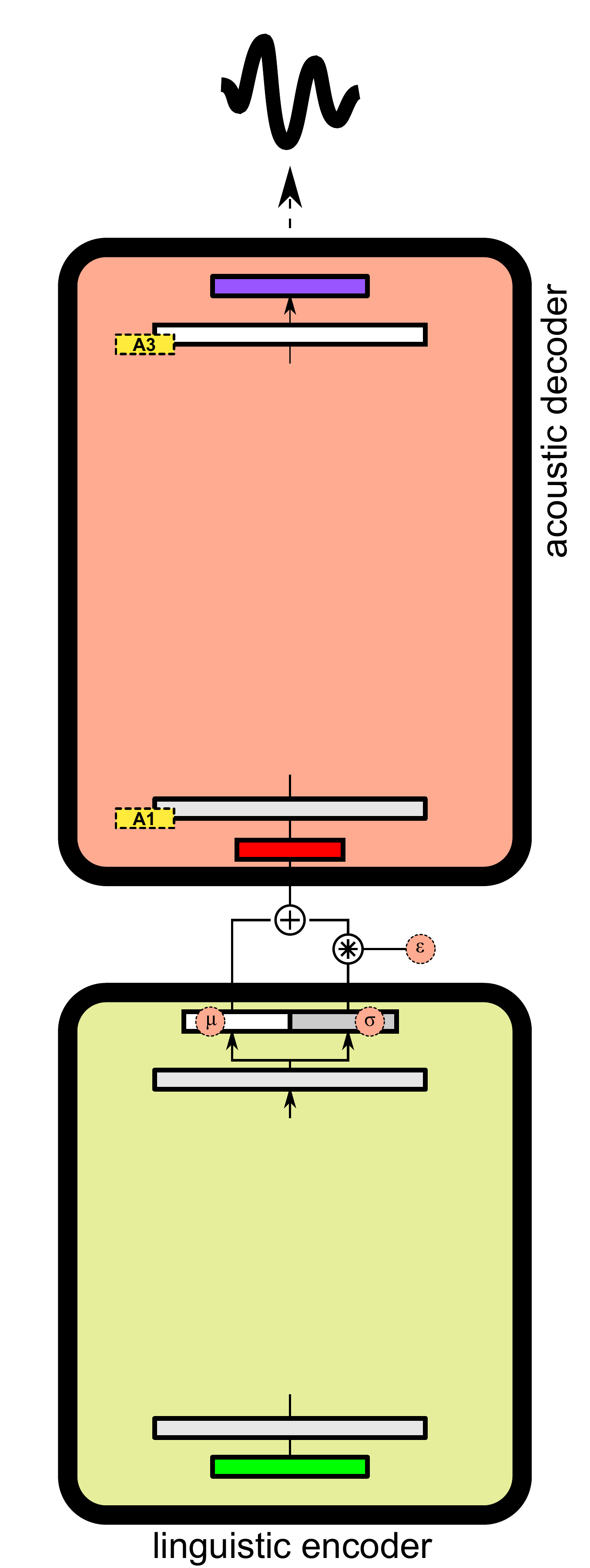}}
  \hfil
  \subfloat[Supervised][Supervised adaptation\\(Text-to-Speech stack)\label{fig:stage-supervised}]{\includegraphics[height=0.48\linewidth]{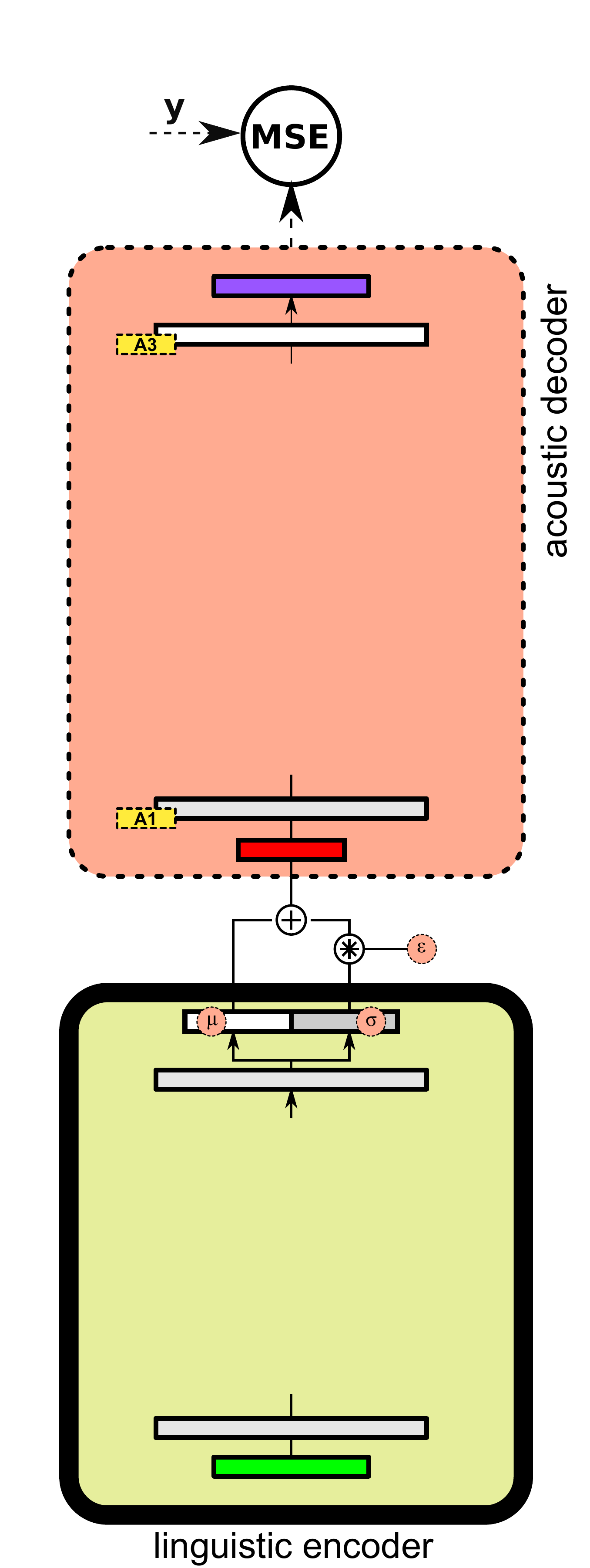}}
  \hfil
  \subfloat[Unsupervised][Unsupervised adaptation\\(Speech-to-Speech stack)\label{fig:stage-unsupervised}]{\includegraphics[height=0.48\linewidth]{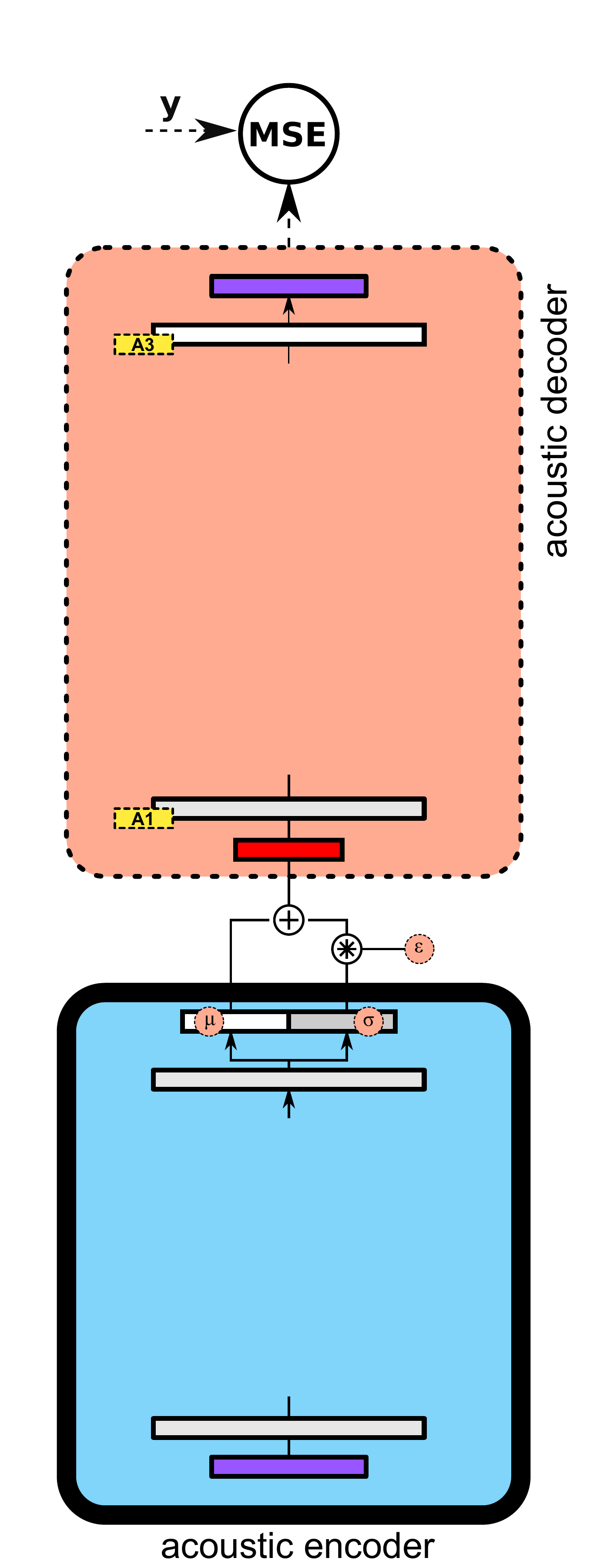}}
\caption{Different modes of the multimodal speaker-adaptive acoustic architecture. Dashed border indicates modules with trainable parameters while bold solid border indicated modules with immutable parameters.}
\label{fig:stages}
\end{figure*}

\subsection{Kullback-–Leibler divergence bound multimodal speech synthesis system}
A conventional acoustic model is a function which transforms linguistic features $\boldsymbol{x} \in \mathbb{R}^{D_x}$ into acoustic features $\boldsymbol{y} \in \mathbb{R}^{D_y}$.
Given the LLE $\boldsymbol{z} \in \mathbb{R}^{D_z}$, the acoustic decoder is a transform function $Dec()$ that maps $\boldsymbol{z}$ to $\boldsymbol{y}$.
$Dec()$ is defined by its parameters $\theta^{\textrm{core}}$ and $\theta^{\textrm{spk},(k)}$, which are the speaker-independent and speaker-dependent parameters, respectively. 

In the training stage, $\theta^{\textrm{core}}$ is trained to focus on the common mapping between the linguistic information and the acoustic output, shared among all speakers, while $\theta^{\textrm{spk},(k)}$ is trained to focus on the unique characteristic of each training speaker. $\theta^{\textrm{spk},(k)}$ can be a speaker scaling, bias or any of speaker components discussed in Section \ref{subsec:proposal-scaling-bias}. Our model assumes that $\boldsymbol{z}$ contains no information about the speaker so the acoustic decoder has to depend on $\theta^{\textrm{spk},(k)}$ in order to reconstruct the speaker characteristics for the acoustic feature output:
\begin{equation}
    \boldsymbol{\tilde{y}} \sim Dec(\boldsymbol{z};\theta^{\text{core}},\theta^{\text{spk},(k)}) = p(\boldsymbol{y}|\boldsymbol{z})
\end{equation}

The linguistic encoder $LEnc()$ encodes a deterministic linguistic feature $\boldsymbol{x}$ to the continuous latent representation $\boldsymbol{z}$. The linguistic encoder is a neural network structure defined by its parameter $\phi^{L}$. To imbue a continuous nature to the latent space of $\boldsymbol{z}$, the output of the linguistic encoder is modeled with a location-scaled distribution inspired by the VAE network. By stacking the linguistic encoder and acoustic decoder, we obtain a complete TTS network with a deterministic linguistic input $\boldsymbol{x}$ and a target acoustic output $\boldsymbol{y}$\footnote{In our implementation, the acoustic decoder only outputs the mean value instead of the density function to simplify the setup.}:
\begin{align}
    \boldsymbol{z}^L &\sim LEnc(\boldsymbol{x};\phi^L) = p(\boldsymbol{z}|\boldsymbol{x})\\
    \boldsymbol{\tilde{y}}^L &= Dec(\boldsymbol{z}^L;\theta^{\text{core}},\theta^{\text{spk},(k)})
\end{align}

The TTS stack can be trained with backpropagation by minimizing the mean square error between the network output $\boldsymbol{\tilde{y}}^L$ and the target acoustic feature $\boldsymbol{y}$%
\begin{equation}
    L_{MSE}(\boldsymbol{\tilde{y}},\boldsymbol{y}) = \frac{1}{D_y}\sum_{i=1}^{D_y}(\tilde{y}_{i}-y_{i})^2 
\end{equation}

When the adaptation data does not include a transcript, we use the acoustic encoder as a substitute for the linguistic encoder. The acoustic encoder $AEnc()$ is a function which transforms the acoustic feature $\boldsymbol{y}$ into the latent variable $\boldsymbol{z}$ by stripping unnecessary information (i.e speaker characteristics) and retaining linguistic information.
The latent output of the acoustic encoder may be used by the acoustic decoder to reconstruct the acoustic feature as follows:
\begin{align}
    \boldsymbol{z}^A &\sim AEnc(\boldsymbol{y};\phi^A) = q(\boldsymbol{z}|\boldsymbol{y})\\
    \boldsymbol{\tilde{y}}^A &= Dec(\boldsymbol{z}^A;\theta^{\text{core}},\theta^{\text{spk},(k)})
\end{align}
We refer the combined network of the acoustic encoder and acoustic decoder as a speech-to-speech (STS) stack. The STS stack is used to adapt the acoustic decoder to unseen speakers when the adaptation data is untranscribed speech.

The key challenge of our proposal is to train a latent variable $\boldsymbol{z}$ that satisfies all of the assumptions made. Previously \cite{luong2018multimodal}, we introduced the joint-goal and tied-layers training methods for this purpose and obtained promising results.
In this paper, we modify the tied-layers training method for the enhanced architecture. More precisely, instead of using cosine or Euler distance functions to measure the difference between two latent samples $\boldsymbol{z}^L$ and $\boldsymbol{z}^A$, we use the Kullback-Leibler (KL) divergence to measure the information lost when using a density function of $\boldsymbol{z}^A$ to approximate the density function of $\boldsymbol{z}^L$. By using the Gaussian as the probability density function we can calculate the KL divergence between two in closed form \cite{ping2018clarinet}\footnote{In our implementation, since we further assume gaussian having diagonal covariance matrix similar to VAE \cite{larsen2015autoencoding}, we calculate KLD of each element of $\boldsymbol{z}$ independently, take the average and then use it as the loss function:
\begin{equation}
    L_{KLD}(\boldsymbol{p},\boldsymbol{q}) = \frac{1}{D_z}\sum_{i=1}^{D_z}KLD(p_{i}, q_{i})
\end{equation}
}.
Modeling LLE as a latent variable and using KL divergence as the tied-layer loss are the two most important modifications to our previous publication \cite{luong2018multimodal}.

\subsection{Different modes of the speaker-adaptive multimodal architecture}

The purpose of the multimodal architecture is that we can use the model in different modes to solve different problems at hand. For speaker-adaptive speech synthesis we need four main modes as illustrated in Fig.\ref{fig:stages}: training, inference, supervised adaptation and unsupervised adaptation.

\subsubsection{Training}
This is the initial mode of the model, in which we need to jointly train all modules to learn a good representation to perform the tasks involved. We used the tied-layers training method, proposed previous in \cite{luong2018multimodal}, to optimize all parameters ($\phi^L$, $\phi^A$, $\theta^{\text{core}}$ and $\theta^{\text{spk},(k)}$ of every training speaker) by minimizing a loss:
\begin{equation}
\label{eq:losstrain}
    \textrm{loss}_{train} = \textrm{loss}_{main} + \beta \; \textrm{loss}_{tie}
\end{equation}
where $\textrm{loss}_{main}$ is the TTS loss calculated as the distortion between the output of the TTS stack and the target acoustic features:
\begin{equation}
    \textrm{loss}_{main} = L_{MSE}(\boldsymbol{\tilde{y}}^L,\boldsymbol{y})
\end{equation}
and the tied-layer loss is the KL divergence between the output of linguistic encoder and that of the acoustic encoder:
\begin{equation}
    \textrm{loss}_{tie} = L_{KLD}(LEnc(\boldsymbol{x}), AEnc(\boldsymbol{y}))
\end{equation}
With this setup, the linguistic encoder and acoustic decoder are trained with a typical TTS acoustic model objective while the acoustic encoder is trained to approximate the linguistic encoder so it could be used as a substitute. By combining the loss and jointly training all modules, we encourage the network to find the optimal representation for all criteria.

\begin{table*}[t]
    \caption{Japanese speech corpus used in the experiments.} 
    \centering
    \scalebox{1.0}{
    \begin{tabular}{|l|r|r|r|r|r|r|r|r|r|}
        \hline
             \multirow{2}{*}{Set} & \multicolumn{2}{c|}{Train (Number of utterances)} & \multicolumn{2}{c|}{Valid (Number of utterances)} & \multicolumn{2}{c|}{Test (Number of utterances)} & \multicolumn{3}{c|}{Number of speakers}\\ \cline{2-10}
           & Each speaker & Total & Each speaker & Total & Each speaker & Total &Male & Female & Total  \\ \hline \hline
         jp.base & $\sim$148 & 34713 & 3 & 705 & - & - & 51 &	184 & 235 \\ \hline
         jp.target.5 & 5 & 100 & \multirow{3}{*}{3} & \multirow{3}{*}{60} & \multirow{3}{*}{10} & \multirow{3}{*}{200} & \multirow{3}{*}{10} & \multirow{3}{*}{10} & \multirow{3}{*}{20}\\
         jp.target.25 & 25 & 500 & & & & & & & \\
         jp.target.100 & 100 & 2000 & & & & & & & \\ \hline

    \end{tabular}}
    \label{tab:dataset}
  \vspace{-1mm}
\end{table*}

\subsubsection{Inference}
As our main task is speech synthesis, in inference mode, we utilize the TTS stack to transform linguistic features into acoustic features with the voice of the desired speaker by using the corresponding speaker component $\theta^{\text{spk},(k)}$.
As a side note, we should confirm that when using the STS stack for inference, the model acts as a many-to-many voice conversion system; we will leave such an investigation for future work.

\subsubsection{Supervised adaptation}
We perform speaker adaptation when we want the model to able to generate speech in the voice of the $r$-th unseen speaker. When both the speech and transcript are available we can adapt the model by using the TTS stack to optimize parameters of the acoustic decoder, as illustrated in Fig.\ \ref{fig:stage-supervised}. In the case of adapting only the speaker components as described in Section \ref{subsec:proposal-scaling-bias}, we train a new set of $\theta^{\text{spk},(r)}$ for $r$-th unseen speaker while keeping the other parameters unchanged. In the case of fine-tuning the entire acoustic decoder as described in Section \ref{subsec:method-finetune}, we first remove all speaker components from the acoustic decoder and then fine tune the remaining parameters $\theta^{\text{core},(r)}$ to the target speaker. In either case, the adapted parameters are obtained by minimizing the distortion between the output of the TTS stack and the natural features:
\begin{equation}
\label{eq:supadapt}
    \textrm{loss}_{adapt} = L_{MSE}(\boldsymbol{\tilde{y}}^L,\boldsymbol{y})
\end{equation}
As supervised adaptation and the inference modes use the same TTS stack, it is expected to perform better than unsupervised adaptation.

\subsubsection{Unsupervised adaptation}
When a transcript does not exist, we can perform unsupervised adaptation. The unsupervised adaptation is conducted in a similar manner as the supervised one, but with one difference: the acoustic encoder is used as a substitute for the linguistic encoder, so we do not have to rely on text. The parameters of the acoustic decoder are optimized to minimize the distortion between the output of the STS stack and the natural features:
\begin{equation}
    \textrm{loss}_{adapt} = L_{MSE}(\boldsymbol{\tilde{y}}^A,\boldsymbol{y})
\end{equation}
As we optimize the acoustic decoder using the STS stack in unsupervised adaptation mode but then use the TTS stack in inference mode, this creates a mismatch between adapting and inferring.

\section{Experimental Conditions}
\label{sec:expconditions}

\begin{figure}[tb]
  \centering
  \includegraphics[width=1.0\columnwidth]{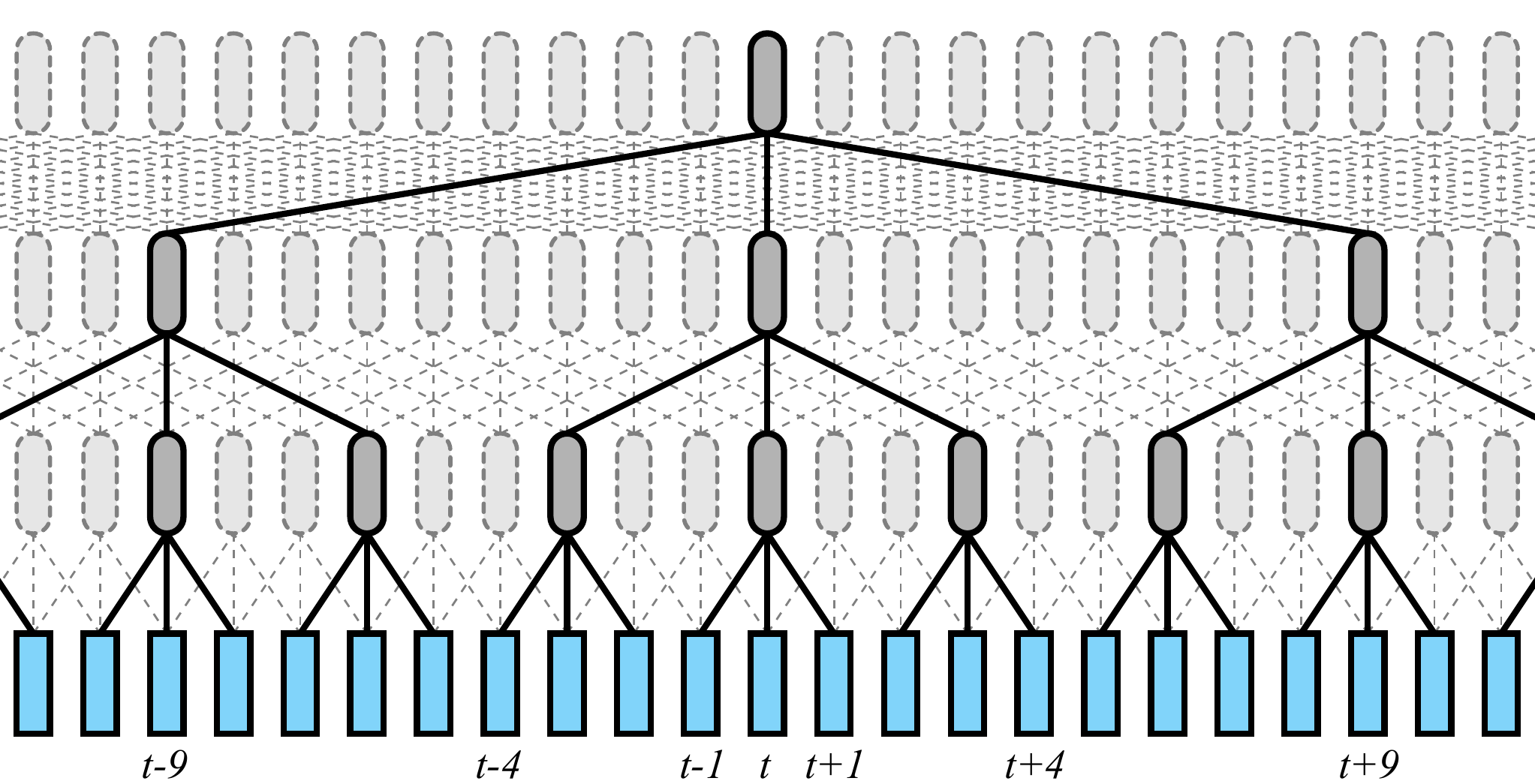}
\vspace{-6mm}
\caption{Temporal contexts captured using a stack of non-overlapping dilated convolution layers.}
\label{fig:tdnn}

\end{figure}

\subsection{Datasets}

We used an in-house multi-speaker Japanese dataset to train the initial multi-speaker model and to conduct speaker adaptation. Table \ref{tab:dataset} shows the details of the data usage.
The setup is similar to our previous study on scaling and bias codes \cite{luong2018scaling} with a slight adjustment to the amount of data used for adaptation.
The objective results are calculated on 200 utterances from twenty speakers included in the jp.target test set. One should note that the data used to train the initial model jp.base is gender-imbalanced with more female speakers than male.

\subsection{Acoustic model configuration}

\begin{figure}[tb]
  \centering
  \subfloat[Linguistic encoder\label{fig:gate-linguistic}]{\includegraphics[width=0.41\columnwidth]{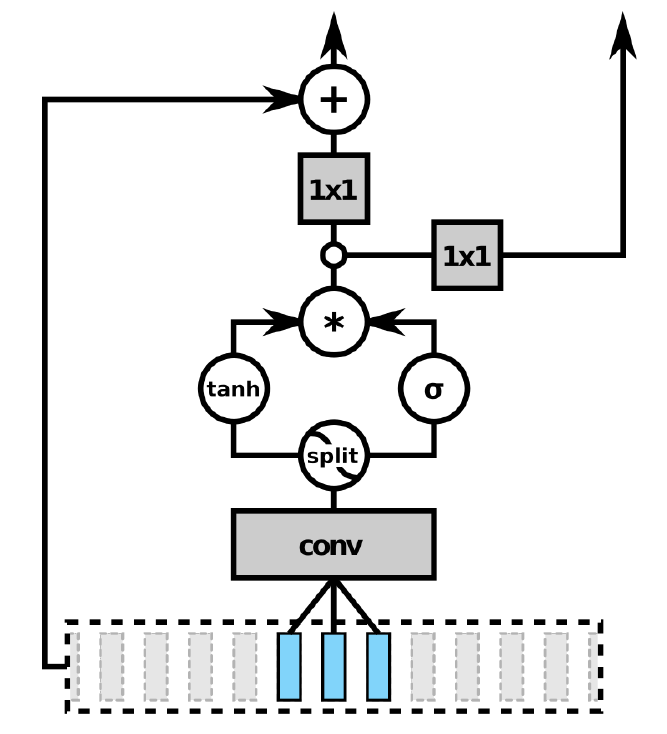}}
  \hfil
  \subfloat[Acoustic en(de)coder\label{fig:gate-acoustic}]{\includegraphics[width=0.41\columnwidth]{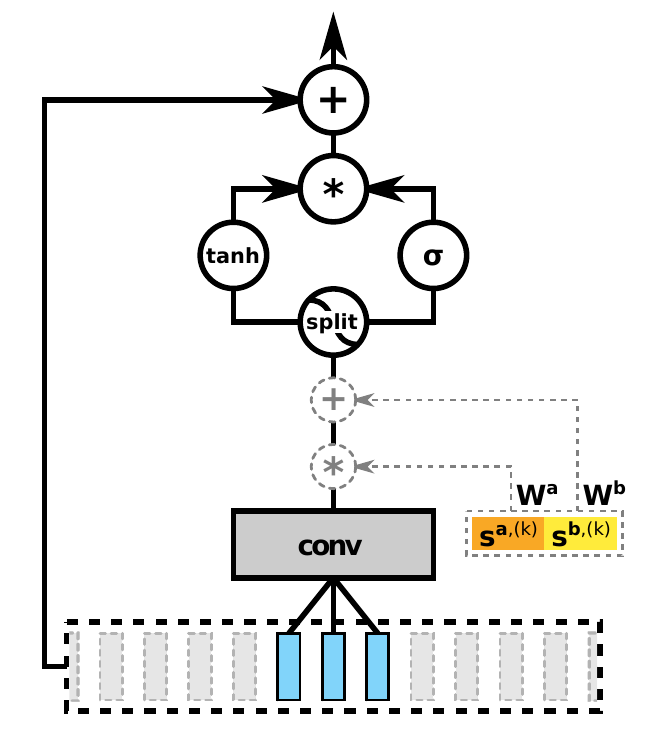}}
\caption{Gated units of convolution layers used in the experiments.}
\label{fig:gated}
\end{figure}

Our acoustic networks contain two types of layer: feedforward and dilated convolution.
$tanh$ is the activation function of most feedforward layers, but the last hidden layer and the output layer of the acoustic decoder use a linear function instead.
The dilated convolution layer is a variation of a time delay neural network (TDNN) \cite{waibel1989phoneme,peddinti2015time}. The convolution layer we used is similar to the dilated causal convolution layer used in the WaveNet model \cite{van2016wavenet} without causality being enforced.
We use the blocks of dilated convolution layers to capture both left and right non-overlapping contexts by setting the dilation rate in order of 1, 3, 9 and 27 as illustrated in Fig.\ \ref{fig:tdnn}.
Two types of gated unit are used with the convolution layer. The first type (Fig.\ \ref{fig:gate-linguistic}) is used in the linguistic encoder and has a residual and a skip output, with trainable weights for both. The second type (Fig.\ \ref{fig:gate-acoustic}) is used in acoustic encoder and acoustic decoder and only has a residual output. Optionally the layers of the acoustic decoder can contain speaker components like speaker scaling and bias, as defined in Equation \ref{eq:speakergate}.

Fig.\ \ref{fig:architecture} is the blueprint of our configuration; the structure of the network is designed to be representative and convenient for testing our hypothesis.
Each module is a standalone network and share a similar structure. The input is transformed to a higher representation with two nonlinear hidden layers. The subsequent block(s) of convolution layers are used to capture temporal context. One last hidden layer is added before the hidden representation is transformed to the desired output.
In the case of the linguistic and acoustic encoders, the output is a density function of LLE; therefore given latent representation of the last hidden layer $\boldsymbol{h}_z$, the output layers of the encoders transform $\boldsymbol{h}_z$ into the mean and standard deviation of the density function of $\boldsymbol{z}$.
\begin{align}
    \boldsymbol{\mu} &= \boldsymbol{W}^\mu \boldsymbol{h}_z \\
    \boldsymbol{\sigma} &= \operatorname{exp}(\boldsymbol{W}^\sigma \boldsymbol{h}_z)
\end{align}
The exponential function $\exp$ is used to make sure the standard deviation will always receive a positive value.
We then apply the reparameterization trick to make the network differentiable:
\begin{equation}
    \boldsymbol{z} = \boldsymbol{\mu} + \boldsymbol{\sigma} \odot \boldsymbol{\epsilon},\quad \boldsymbol{\epsilon} \sim \mathcal{N}(0,1)
\end{equation}
All hidden layers of the linguistic and acoustic encoders have 128 units while the hidden layers of the acoustic decoder have 256 unit. LLE was set to be a 64-dimensional feature.

\subsection{WaveNet vocoder configuration}
\label{subsec:wavenetcfg}
We trained a speaker-independent WaveNet vocoder \cite{hayashi2017investigation} using the jp.base training set. We then used the model to generate speech for the target unseen speakers. While fine-tuning the WaveNet vocoder \cite{liu2018wavenet,sisman2018voice} to the target unseen speaker is reported to improve performance, we decided to use a single SI WaveNet vocoder in all of the experiments and focus on evaluation the performance of the acoustic model. WaveNet is trained to model a 16kHz waveform which is quantified into a 10-bit u-law. The network contains 40 dilated causal layers conditioned on a mel-spectrogram. We kept the setup for WaveNet simple and similar to that of the original study \cite{van2016wavenet}.

\subsection{Feature pre-processing}
\subsubsection{Linguistic features}: We used standard linguistic features of Japanese speech synthesis. The features contained quinphone contexts, word part-of-speech tags, pitch accent types of the accent phrases, interrogative phrase marks, and other structural information such as the position of the mora in a word, accent phrases, and utterances. We aligned the linguistic features with the acoustic features by using an external system \cite{luong2017adapting}. The linguistic features were then concatenated with duration information into a 389-dimensional vector. As we are interested in adaptation of acoustic models, we did not train or adapt the duration model and instead used the oracle duration obtained from the reference utterances to generate the synthetic speech.

\subsubsection{Acoustic features}
We simplified our setup by using an 80--dimensional mel-spectrogram as the acoustic feature compared to our previous study \cite{luong2018scaling} where we used multiple types of vocoder features. The features are extracted from a 25ms window and shifted in steps of 5ms over the speech waveform. The WaveNet vocoder is used to synthesize speech waveform from the mel-spectrogram feature. With this setup we could use a mean square error metric for both the training loss and the objective evaluation.
For the objective evaluation, we removed silence frames, indicated by the linguistic features, before calculating the mean square error in order to obtain results more focused on speech regions.

\begin{table}[tb]
    \caption{Objective evaluation of the baseline system for multi-speaker and supervised adaptation tasks under various strategies.}
    \centering
    \scalebox{1.0}{
    \begin{tabular}{llrrrr}
        \hline \hline
        \multirow{2}{*}{Strategy} & \multirow{2}{*}{Layer} & \multirow{2}{*}{Speaker bias} & \multicolumn{3}{c}{Mean square error} \\
        & & & 5 utts & 25 utts & 100 utts \\ \hline
        \texttt{MU-A1b} & A1 & 128 & 0.555 & - & 0.533 \\
        \texttt{MU-A1B} & A1 & full & 0.553 & - & 0.518 \\ \hline
        \texttt{AD-A1b} & A1 & 128 & 0.584 & 0.560 & 0.553 \\
        \texttt{AD-A1bB} & A1 & 256 & 0.578 & 0.555 & 0.545 \\
        \texttt{AD-A1B} & A1 & full & 0.578 & 0.553 & 0.544 \\ \hline
    \end{tabular}}
    \label{tab:baseline}
\end{table}

\subsection{Training and adapting optimization}

The initial multi-speaker models were trained to minimize the designate loss in Equation \ref{eq:losstrain}, with the tied-layers factor $\beta$ set to 0.25 for all strategies.
The training stopped naturally after 5 epochs without any improvement on the validate set or was forcefully stopped at the 128th epoch, the last best epoch is used as the final model. In practice, all training is naturally stopped by about the thirtieth epoch.
The speaker adaptation followed a similar scheme but only a certain part of the acoustic decoder was optimized instead of the entire network. For the adaptation with five utterances, many strategies were forcefully stopped at the 128th epoch. For the adaptations with 25 and 100 utterances, the adaptation usually converged after 5 epochs without further improvements. 

\begin{figure*}
  \centering
  \includegraphics[width=0.95\linewidth]{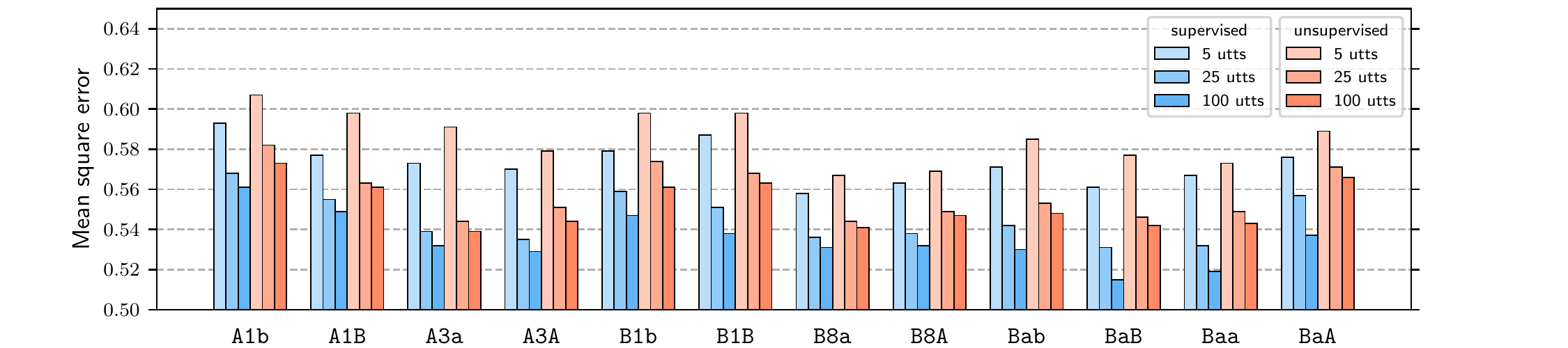}
  \vspace{-3mm}
\caption{Objective evaluations of supervised and unsupervised speaker adaptation of multiple strategies utilizing speaker scaling and bias.}
\vspace{-3mm}
\label{fig:objective}
\end{figure*}

\section{Evaluation and Discussion}
\label{sec:evaluations}

\subsection{Baseline objective evaluations of the conventional multi-speaker and supervised adaptation tasks}
\label{subsec:eval-baseline}

We first evaluate several baselines without the proposed elements.
These baseline systems only contain the linguistic encoder and the acoustic decoder, where the linguistic encoder outputs a deterministic latent variable $\boldsymbol{z}$ instead of a density function. 
We add either a bias code $\boldsymbol{s}^{b,(k)}$ or speaker bias $\boldsymbol{b}^{(k)}$ to the A1 layer (Fig.\ \ref{fig:architecture}) to represent the conventional speaker codes approach.
The strategies we investigate for baseline are shown in Table \ref{tab:baseline} for multi-speaker task and speaker adaptation task.
\texttt{MU} models are trained using the training data of jp.base combined with either jp.target.5 or jp.target.100.
On the other hand, the initial model of the speaker adaptation task \texttt{AD} is trained using training data of jp.base; adaptation is then performed by optimizing the speaker components for unseen speakers with jp.target.\{5,25,100\} data.

From the table, we can first see that there is little difference between two multi-speaker strategies \texttt{MU-A1b} and \texttt{MU-A1B} when the amount of data is limited. However \texttt{MU-A1B} shows a greater improvement when more data become available. Similarly for the speaker adaptation task, there is little difference between \texttt{AD-A1bB} and \texttt{AD-A1B}, and both shown slightly better performance than \texttt{AD-A1b}.
The multi-speaker \texttt{MU-A1B} strategy is better than the speaker adaptation \texttt{AD-A1B} counterpart but speaker adaptation is faster and more convenient to conduct: the same conclusion was reached in our previous studies \cite{luong2017adapting,luong2018scaling}.

\subsection{Preliminary objective evaluations on supervised and unsupervised adaptation with speaker scaling and bias}

\begin{table}[tb]
    \caption{Adaptation strategies utilizing speaker scaling and bias}
    \centering
    \scalebox{1.0}{
    \begin{tabular}{llrrr}
        \hline \hline
        \multirow{2}{*}{Strategy} & \multirow{2}{*}{Layer(s)} & \multicolumn{3}{c}{Number of speaker parameters} \\
         & & Speaker bias & Speaker scale & Total \\ \hline
        \texttt{A1b} & A1 & 128 & 0 & 128x1 \\
        \texttt{A1B} & A1 & full & 0 & 256x1 \\
        \texttt{A3a} & A3 & 128 & 128 & 256x1 \\
        \texttt{A3A} & A3 & full & full & 512x1 \\ \hline
        \texttt{B1b} & B1 & 128 & 0 & 128x1 \\
        \texttt{B1B} & B1 & full & 0 & 512x1 \\
        \texttt{B8a} & B8 & 128 & 128 & 256x1 \\
        \texttt{B8A} & B8 & full & full & 1024x1 \\ \hline
        \texttt{Bab} & B[1-8] & 64 & 0 & 64x8 \\
        \texttt{BaB} & B[1-8] & full & 0 & 512x8 \\
        \texttt{Baa} & B[1-8] & 64 & 64 & 128x8 \\
        \texttt{BaA} & B[1-8] & full & full & 1024x8 \\ \hline
    \end{tabular}}
    \label{tab:strategies}
\end{table}

We evaluated our methods described in Section \ref{sec:multimodal} for both supervised and unsupervised speaker adaptation. We also investigated multiple strategies for modeling the speaker transformation, as shown in Table \ref{tab:strategies}.
Here, \texttt{A}-type is a strategy in which the speaker component is injected at a feedforward layer. \texttt{A1} represents the conventional speaker embedding, while \texttt{A3} represents the best strategy in our preliminary study \cite{luong2018scaling} with both scaling and bias codes at a layer near the output.
\texttt{B}-type is a strategy in which the speaker components are injected at the convolution layers. \texttt{Ba} is where all eight convolution layers of the acoustic decoder have their own speaker components.

Figure \ref{fig:objective} shows the results of the objective evaluations of the strategies listed in Table \ref{tab:strategies} for both supervised and unsupervised adaptation tasks.

\subsubsection{Comparison with the baseline}
The supervised adaptation of \texttt{A1b} is slightly worse than \texttt{AD-A1b}, the same goes for \texttt{A1B} and \texttt{AD-A1B}. Comparing between different amounts of data, we conclude that the degradation in performance is an acceptable trade-off for the ability to perform unsupervised speaker adaptation.

\subsubsection{Supervised and unsupervised adaptation}
For the \texttt{A}-type strategies, the unsupervised adaptation consistently improves when the supervised adaptation improves, which validates our method.
Both supervised and unsupervised systems of the best strategy identified so far, \texttt{A3a}, is as good as the supervised adaptation baseline \texttt{AD-A1B}.

\subsubsection{Speaker scaling and bias at residual gated layer}
We trained a couple strategies with speaker scaling and bias and put them in layer \texttt{B1} or \texttt{B8}. Figure \ref{fig:objective} shows that these strategies performed as well as \texttt{A}-type strategies. This confirms that speaker scaling and bias can be used at either feedforward or convolution layers.

\subsubsection{Speaker components at multiple layers}
The $\texttt{Ba}$ strategies have speaker scaling and bias at all eight residual convolution layers. $\texttt{BaB}$ obtained the best results of all those evaluated, without showing any overfitting in the adaptation of five-utterances case.

\begin{figure}[t]
  \subfloat[Female speaker]{\includegraphics[width=0.98\columnwidth]{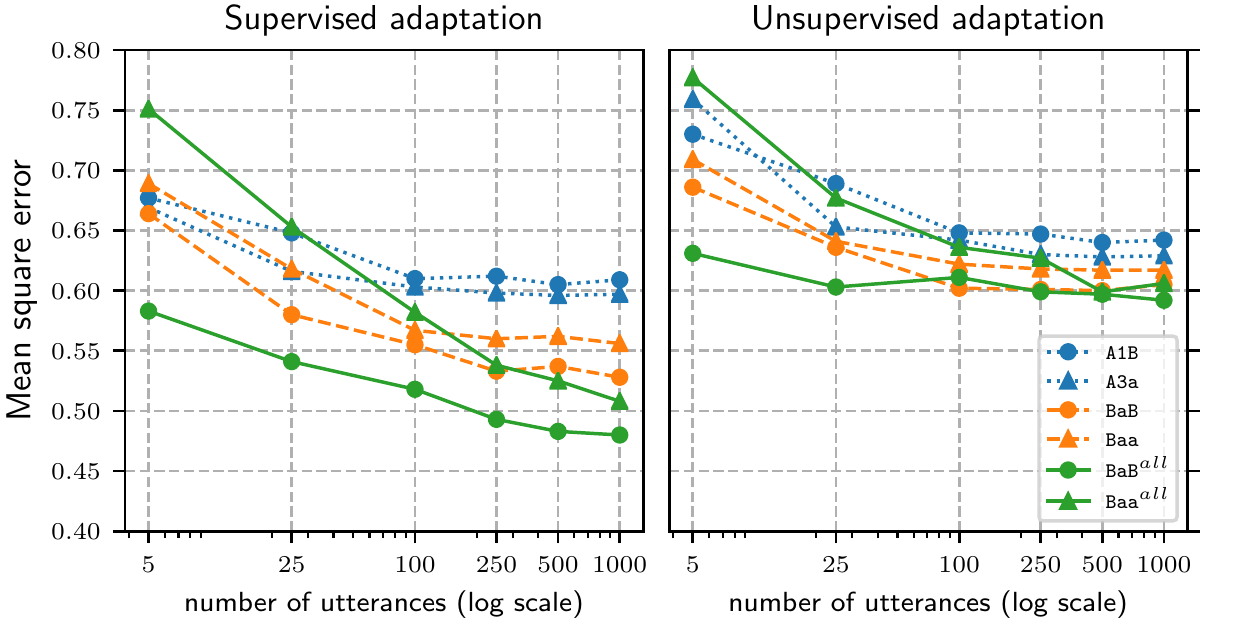}} \\
  \subfloat[Male speaker]{\includegraphics[width=0.98\columnwidth]{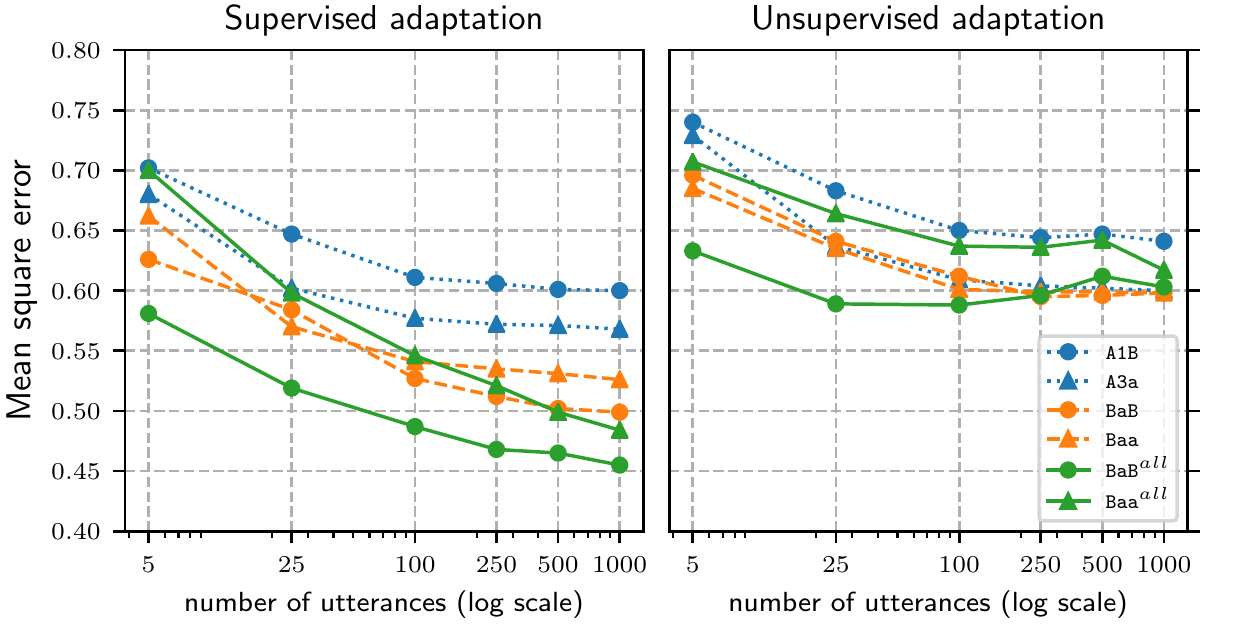}}
  \caption{Objective evaluation of adaptation for male speaker. Mean square error between the natural and generated feature. }
  \label{fig:obj_amount}
  \vspace{-3mm}
\end{figure}

\begin{figure*}
  \subfloat[Female Speaker\label{fig:subjective-female}]{
  \begin{minipage}[c]{0.485\textwidth}
    \includegraphics[height=0.55\linewidth]{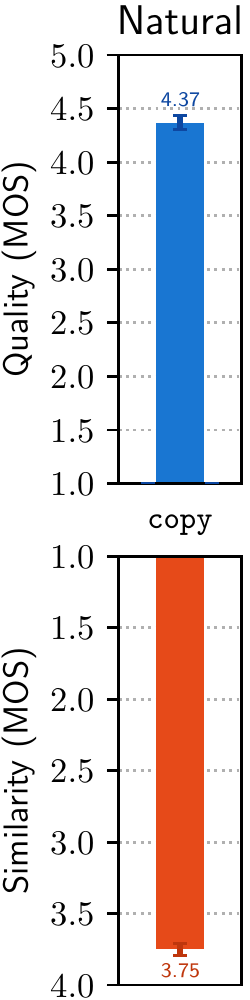}\includegraphics[height=0.55\linewidth]{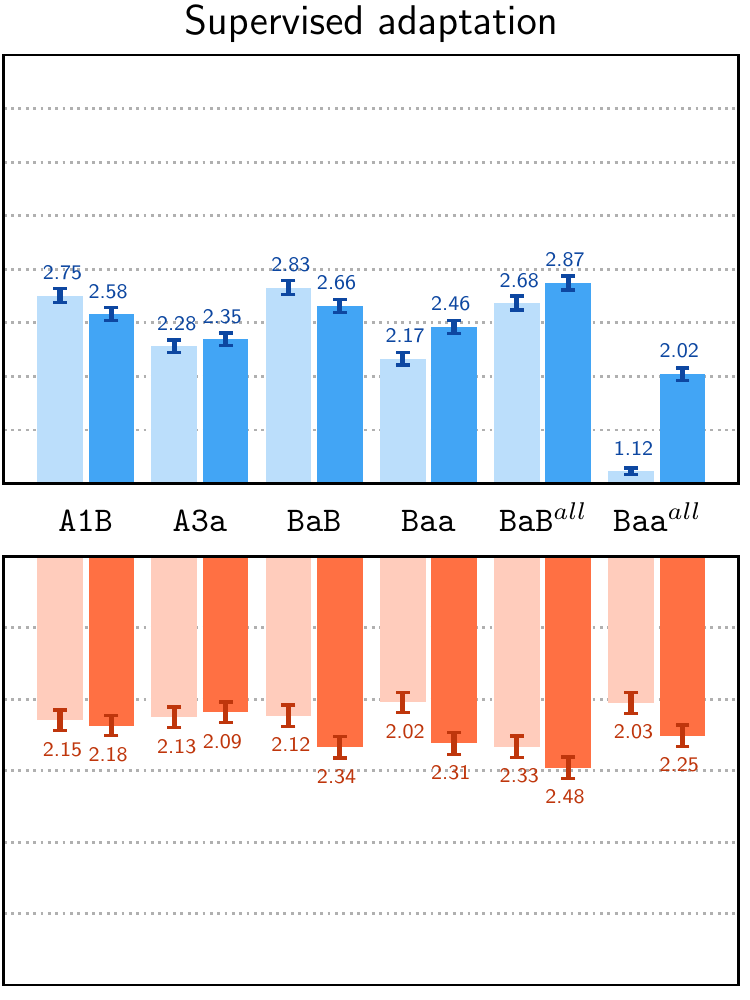}\includegraphics[height=0.55\linewidth]{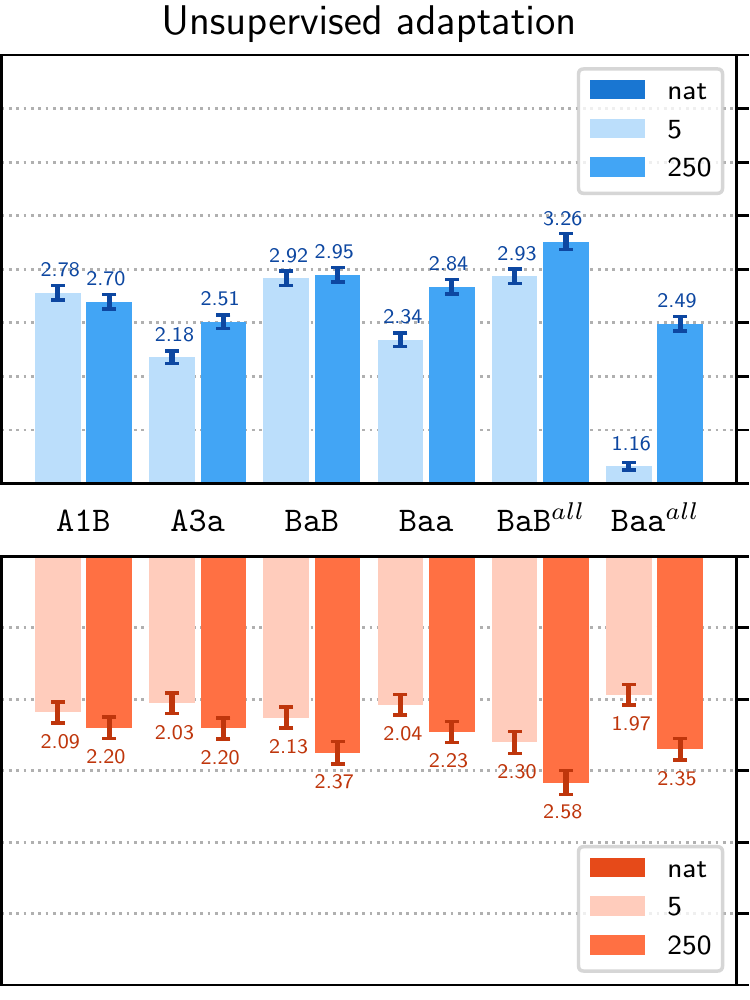}
  \end{minipage}
  }
  \hfill
  \subfloat[Male Speaker\label{fig:subjective-male}]{
  \begin{minipage}[c]{0.485\textwidth}
    \includegraphics[height=0.55\linewidth]{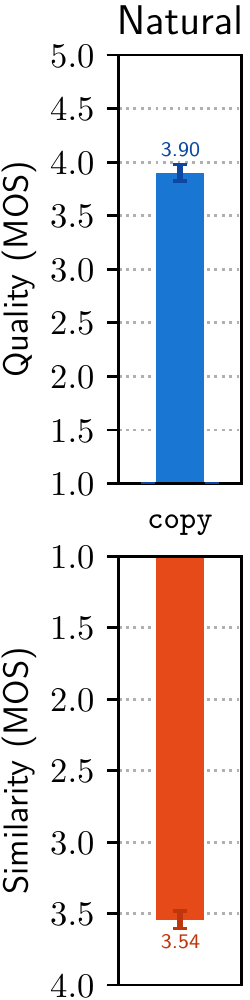}\includegraphics[height=0.55\linewidth]{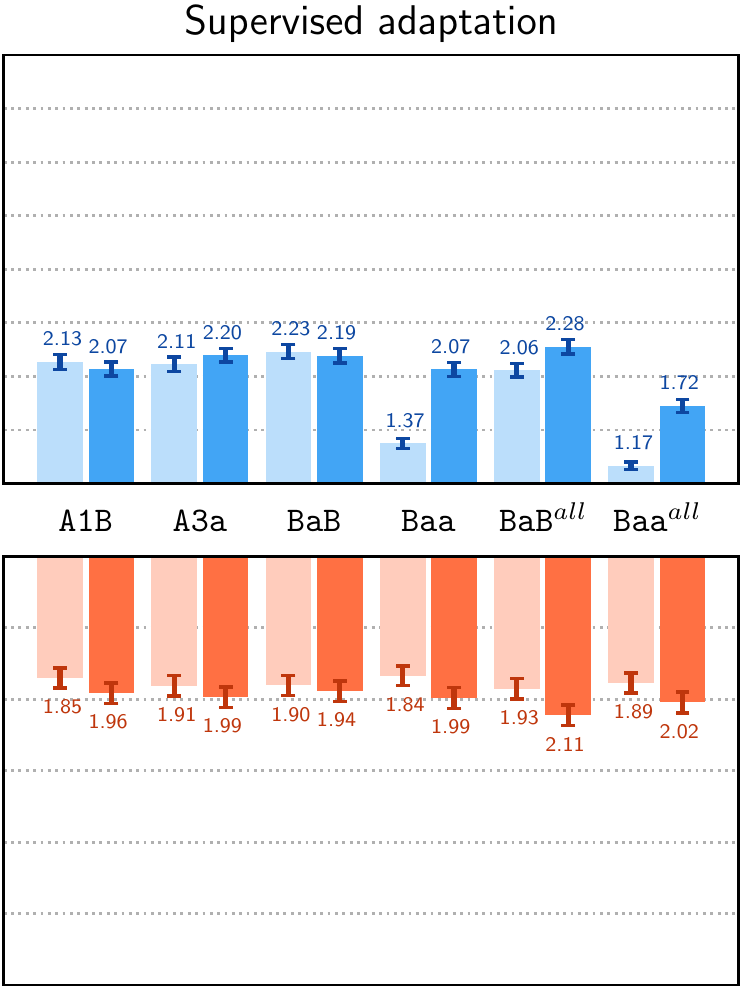}\includegraphics[height=0.55\linewidth]{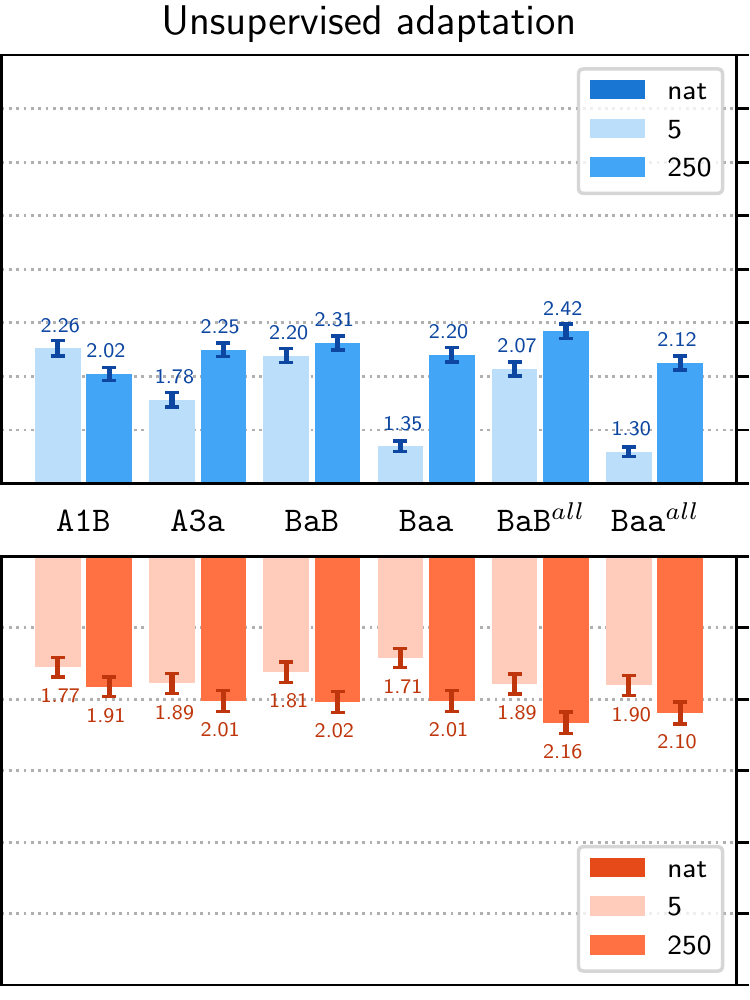}
  \end{minipage}
  }
\caption{Subjective results of quality and similarity test. Samples were generated from six strategies for supervised and unsupervised adaptation using either 5 or 250 utterances. The error bar indicates the 95\% confidence interval.}
\vspace{-3mm}
\label{fig:subjective}
\end{figure*}

\subsection{Focused objective and subjective evaluations and strategies of adapting the entire network}
\label{subsec:focused}
Here, we evaluate the adaptation performance for just two target speakers, 1 male and 1 female, who have more speech data.
The initial models used in previous section are reused to adapt to these two target speakers: \texttt{A1B} is used as the new baseline, \texttt{A3a} is a strategy with both scaling and bias code, while \texttt{BaB} and \texttt{Baa} are those that obtained the best objective evaluations in the previous section. Finally two new strategies $\texttt{BaB}^\texttt{all}$ and $\texttt{Baa}^\texttt{all}$ are introduced for the method described in Section \ref{subsec:method-finetune}. For these strategies, we first remove the speaker components from pretrained \texttt{BaB} and \texttt{Baa} models; then we adapt the remaining parameters of the acoustic decoder to the target speaker.

\subsubsection{Objective evaluation}
The objective evaluations are shown in Fig.\ \ref{fig:obj_amount}; the results are calculated from 100 test utterances of each speaker. 
The number of utterances used for adaptation ranged from 5 to 1000. We can see that \texttt{A3a} is still slightly better than \texttt{A1b} at most data points.
Among the legacy strategies, \texttt{BaB} benefits the most from the increase in data. For the new strategies, $\texttt{BaB}^\texttt{all}$ surprisingly outperforms all others, while $\texttt{Baa}^\texttt{all}$ shows poor results when the amount of data is limited. The pattern is consistent between male and female speakers.
For the unsupervised adaptation task, $\texttt{BaB}^\texttt{all}$ also seems to be the best strategy. However, adding more adaptation data seems to worsen the objective results.
There is still a gap between the performances of the supervised and unsupervised adaptation intra-strategies, but the best proposed strategy $\texttt{BaB}^\texttt{all}$ surpasses the baseline \texttt{A1B} in both supervised and unsupervised adaptation tasks.

\subsubsection{Subjective evaluation}

\begin{figure}[t]
  \centering
  \subfloat[Female speaker]{\includegraphics[width=0.975\columnwidth]{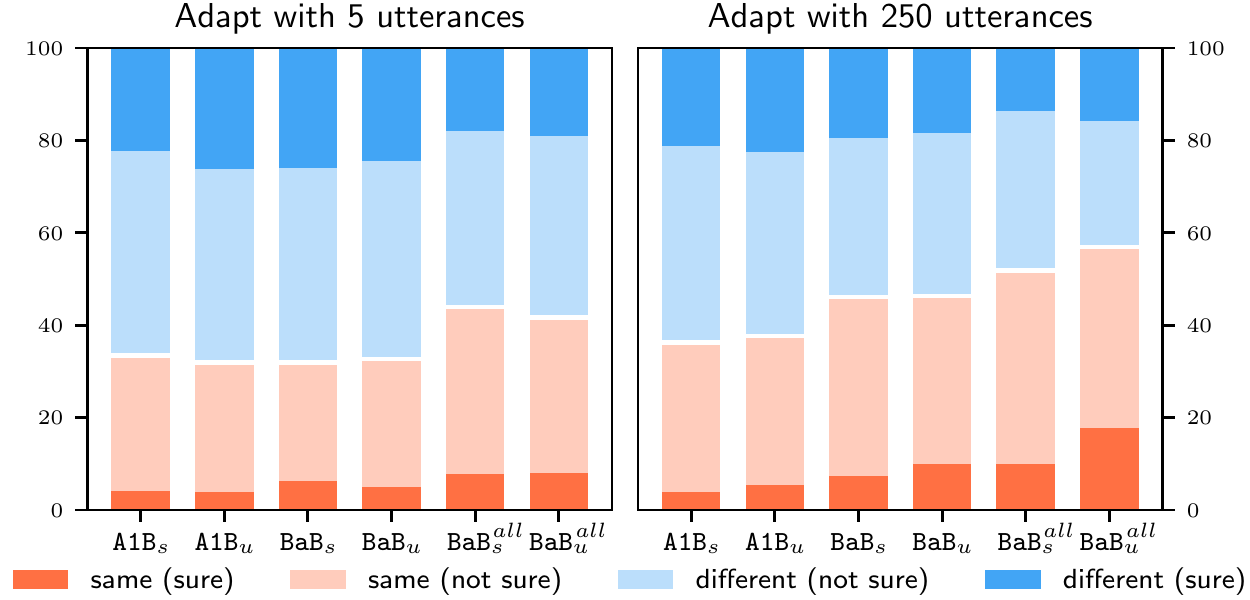}} \\
  \subfloat[Male speaker]{\includegraphics[width=0.975\columnwidth]{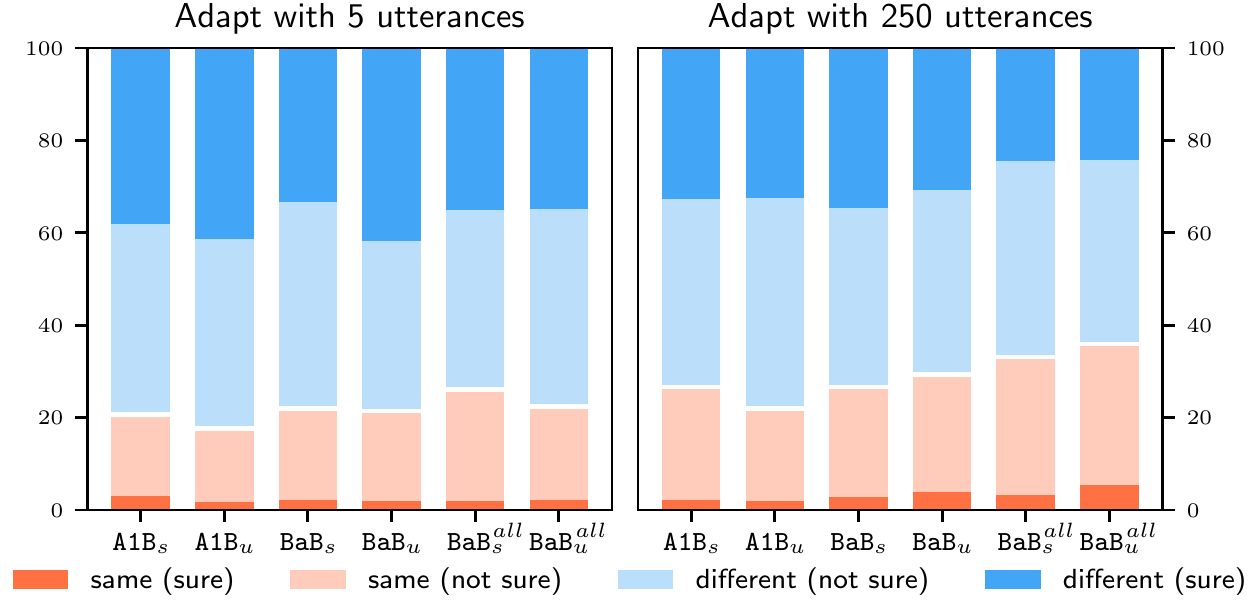}}
  \caption{Detailed results of similarity evaluations for selected strategies. The subscript $_s$ denotes supervised while $_u$ denotes unsupervised adaptation.}
  \label{fig:similarity}
  \vspace{-3mm}
\end{figure}

We conducted subjective surveys on the supervised and unsupervised adaptation tasks. To reduce the number of systems that the participants had to evaluate, we only used models adapted with 5 and 250 utterances\footnote{Speech samples are available online at \url{https://nii-yamagishilab.github.io/sample-tts-unified-adaptation/}}.
The SI WaveNet vocoder was used to synthesize waveforms from the generated mel-spectrogram. A copy synthesis system was also included as a reference.
For the quality test, participants were asked to judge the quality of a sample in terms of a 5-point scale mean opinion score (MOS). For the similarity test, participants were asked to judge the similarity between a generated sample and the recorded sample in a 4-point scale MOS test where 1 means different (sure), 2 different (not sure), 3 the same (not sure) and 4 is the same (sure). One session consisted of 25 quality and 25 similarity questions, with one question for each system. The final results were calculated from only those sessions in which all 50 questions were answered. Each session contained samples of either female or male speakers. We collected in total 500 sets for the female speaker and 497 sets for the male speaker from a total of 198 paid participants, who each did ten sessions at most.

The mean values of the quality and similarity tests are shown in Fig.\ \ref{fig:subjective}. Several inter-speaker and inter-strategie trends are: the quality of the male speaker is lower than the female speaker; when more data becomes available the similarity score increases for most strategies, while the quality sometimes decreases; strategies utilizing both speaker scaling and bias got worse results than those utilizing only speaker bias, despite their better objective results; the supervised and unsupervised adaptations strategies gave similar results.
Generally speaking, $\texttt{BaB}^\texttt{all}$ is the best strategy for both supervised and unsupervised adaptation tasks, especially in the 250-utterance case.
The most surprising outcome is that unsupervised adaptation of $\texttt{BaB}^\texttt{all}$ outperforms its supervised counterpart, even though the objective results indicated the opposite.
Figure \ref{fig:similarity} shows the details of the similarity test of \texttt{A1B}, \texttt{BaB} and $\texttt{BaB}^\texttt{all}$. The unsupervised adaptation of $\texttt{BaB}^\texttt{all}$ using 250 utterances has the most positive results for both male and female speakers.

\subsection{Boosting performance and comparing with baseline systems}

\begin{table*}[t]
    \caption{VCTK English speech corpus used in the experiments.} 
    \centering
    \scalebox{1.0}{
    \begin{tabular}{|l|r|r|r|r|r|r|r|r|r|}
        \hline
             \multirow{2}{*}{Set} & \multicolumn{2}{c|}{Train (Number of utterances)} & \multicolumn{2}{c|}{Valid (Number of utterances)} & \multicolumn{2}{c|}{Test (Number of utterances)} & \multicolumn{3}{c|}{Number of speakers}\\ \cline{2-10}
           & Each speaker & Total & Each speaker & Total & Each speaker & Total &Male & Female & Total  \\ \hline \hline
         en.base & $\sim$367 & 26405 & 5 & 360 & - & - & 31 & 41 & 72 \\ \hline
         en.target.10 & 10 & 80 & \multirow{2}{*}{5} & \multirow{2}{*}{40} & \multirow{2}{*}{15} & \multirow{2}{*}{120} & \multirow{2}{*}{4} & \multirow{2}{*}{4} & \multirow{2}{*}{8}\\
         en.target.320 & 320 & 2560 & & & & & & & \\ \hline

    \end{tabular}}
    \label{tab:vctk-corpus}
  \vspace{-3mm}
\end{table*}

By keeping the difference in the experimental environment at a minimum between strategies, the previous experiment gave us an expectations on the performance of the proposed method in different configurations. 
Next, we reproduced several key strategies with an English multi-speaker corpus and compared them with several baselines.
The VCTK corpus \cite{veaux2017superseded} was used for this experiment. The data partition between the base and target speakers is described in Table \ref{tab:vctk-corpus}. 
New acoustic models and the WaveNet vocoder were trained with this corpus to make the experiment self-contained. All the experimental conditions were the same as described in Section \ref{sec:expconditions} unless stated otherwise. As the majority of speakers had English accents, the linguistic features were based on the Combilex lexicon for received pronunciations \cite{richmond2009robust}.
We used the supervised and unsupervised systems of the $\texttt{BaB}$ and $\texttt{BaB}^\texttt{all}$ strategies as candidates of our method.

\subsubsection{Changes and complementary tactics}
Several new tactics were deployed to boost the quality of the synthetic samples:

\begin{itemize}
  \item The sampling rate of the SI WaveNet vocoder was increased to 24kHz without changing the network configuration. We also fine-tuned it to each individual target speaker with the corresponding amount of adaptation data.
  \item In inference mode, $\boldsymbol{z}^L$ is assigned the mean of the Gaussian instead of samples from it. This should remove any stochasticity from the generating process and reduce the chances of getting extraordinary values.
  \item The subjective results shown in Section \ref{subsec:focused} suggested that unsupervised strategies are better than supervised ones. This phenomenon is caused by the mismatch between the optimized objective (mean square error) and the desired objective (human perception), which is not a simple problem that can be tackled directly. Because of that, we devised a simple intuitive tactic to push the performance of supervised adaptation instead. Specifically, we took the outputs of both TTS and STS stacks into account when adapting with transcribed speech:
  \begin{equation}
  \label{eq:supplusadapt}
    \textrm{loss}_{adapt} = L_{MSE}(\boldsymbol{\tilde{y}}^L,\boldsymbol{y}) + L_{MSE}(\boldsymbol{\tilde{y}}^A,\boldsymbol{y}) 
  \end{equation}
This is essentially the joint-goals method \cite{luong2018multimodal}, but here it is used for the supervised adaptation. We applied this tactic to the supervised systems of $\texttt{BaB}^\texttt{all}$.
\end{itemize}

\begin{figure}[tb]
  \centering
  \includegraphics[width=0.95\columnwidth]{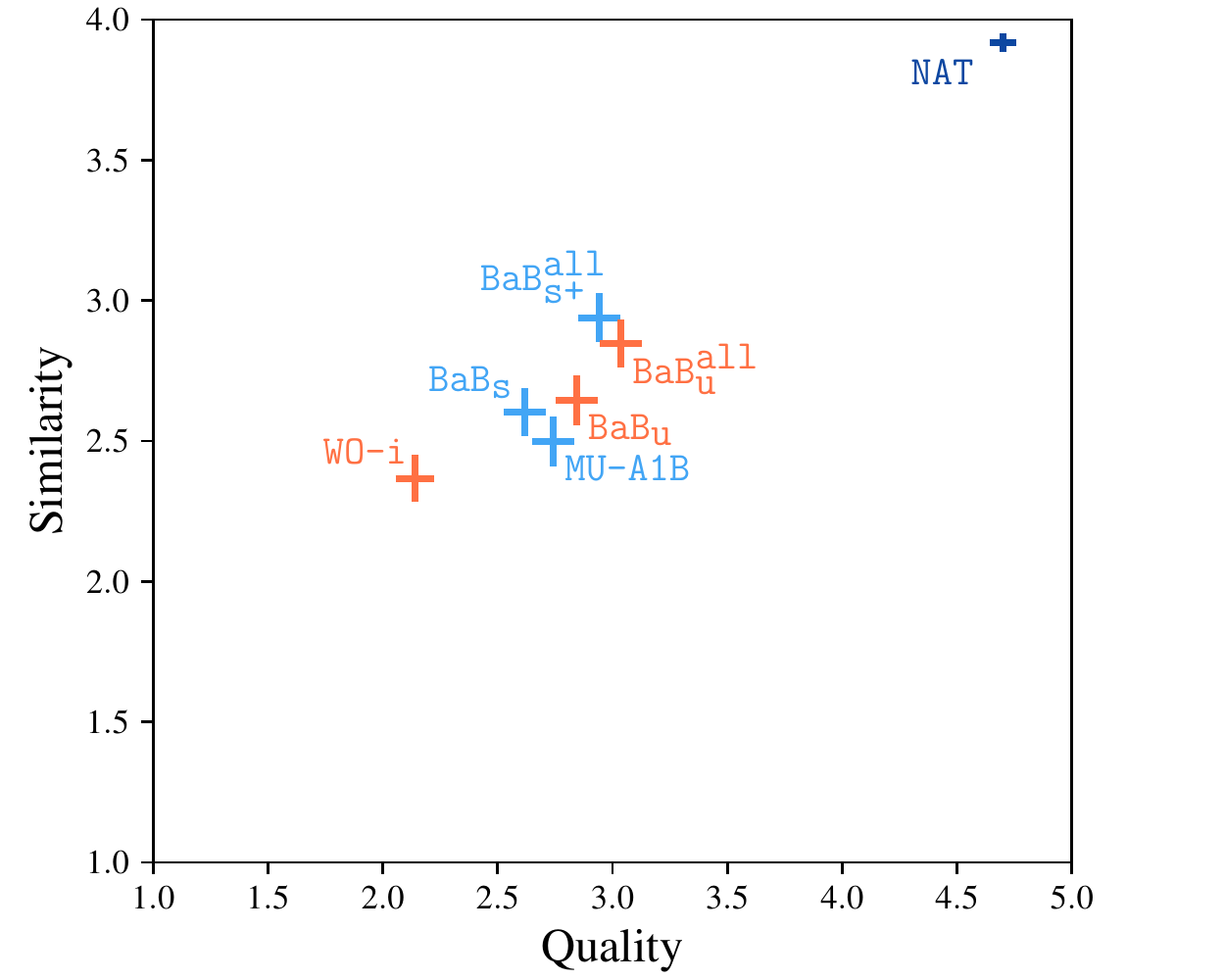}
\vspace{-3mm}
\caption{Subjective results for 10-utterance cases (avg. 33 seconds). The lines indicate 95\% confidence interval.}
\vspace{-3mm}
\label{fig:vctk010}
\end{figure}

\subsubsection{Baseline systems}

We use \texttt{MU-A1B} (described in Table \ref{tab:baseline}) as the supervised baseline. Technically speaking, $\texttt{MU-A1B}$ is not an adaptation strategy but rather a seen speaker scenario which usually sees better performance \cite{jia2018transfer,hojo2018dnn,luong2018scaling}.
\texttt{MU-A1B} shares the same WaveNet vocoder and most of its configuration with the proposed strategies. As the WaveNet vocoder is a trained model and sensitive to the training data \cite{wu2018collapsed,jia2018transfer}, we prepared another baseline with the conventional vocoder as an anchor system.
A speaker-adaptive model using the i-vector \cite{wu2015study,doddipatla2017speaker,takaki2018unsupervised} and WORLD vocoder \cite{morise2016world}, denoted as \texttt{WO-i}, is used as the unsupervised baseline. The acoustic model of \texttt{WO-i} consists of four 1024-unit feedforward layers followed by a 512-unit bi-directional quasi-recurrent layer \cite{bradbury2016quasi} and a linear layer mapping hidden features to the output. While this setup is not the current state-of-the-art, it is a typical one for vocoder based acoustic models \cite{wu2015study,wu2016merlin}. The input of \texttt{WO-i} is the linguistic feature concatenated with a 64-dimensional i-vector represented speaker. The output contains several acoustic features and their dynamic counterpart which would be used to synthesize speech with the WORLD vocoder \cite{wu2016merlin,luong2018scaling}. To extract the i-vector, we trained 1024-gaussian universal background model (UBM) using the MFCC feature extracted from en.base training set. The i-vector of the target speakers was calculated using a corresponding amount of adaptation data. The KALDI toolkit \cite{povey2011kaldi} was used to build the i-vector system.

\begin{figure}[tb]
  \centering
  \includegraphics[width=0.95\columnwidth]{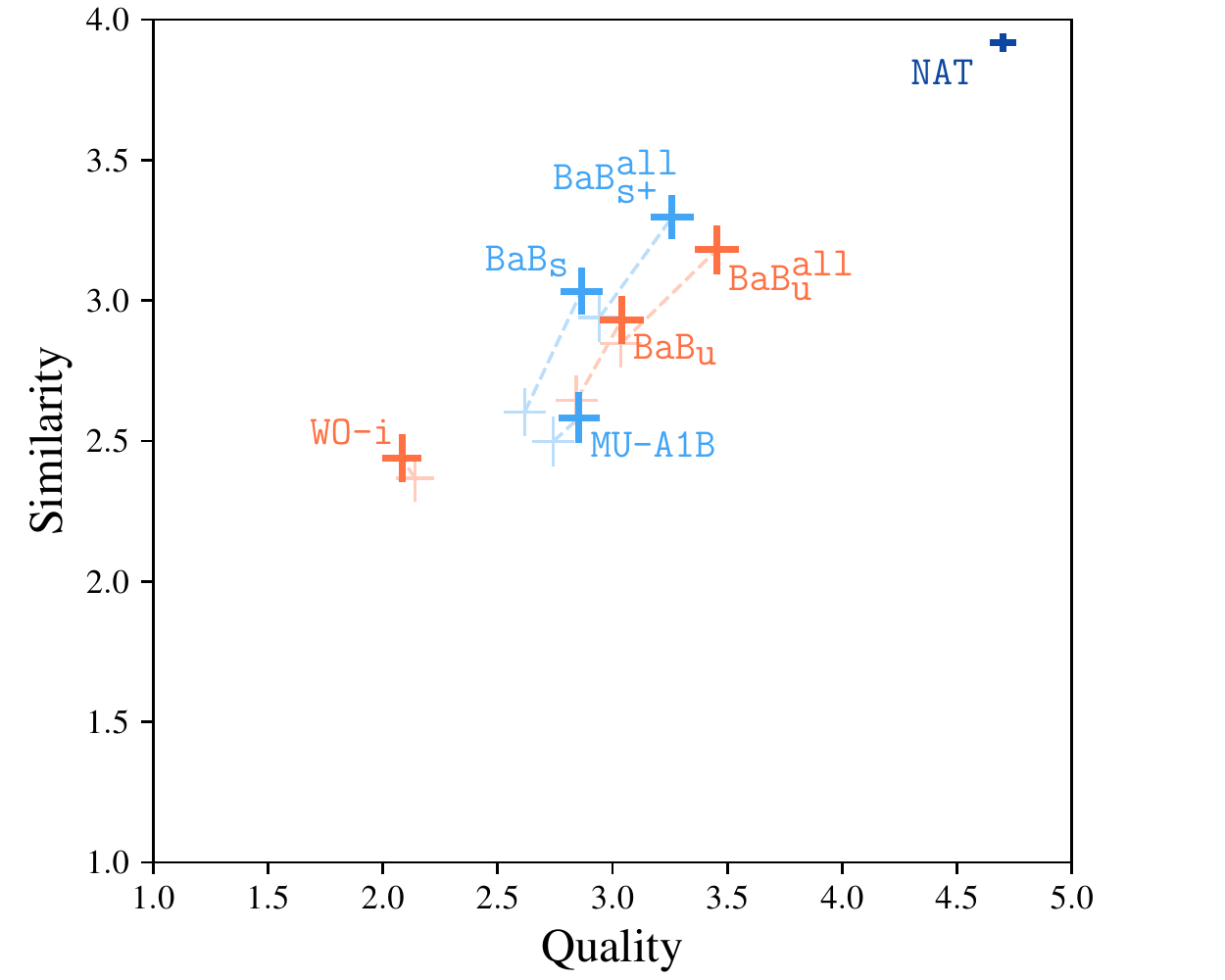}
\vspace{-3mm}
\caption{Subjective results for 320-utterance cases (avg. 17.3 minutes).}
\vspace{-1mm}
\label{fig:vctk320}
\end{figure}

\begin{table*}[t]
    \caption{Details of target speakers in VCTK experiments}
    \centering
    \scalebox{1.0}{
    \begin{tabular}{lrrrrrrrrr}
        \hline \hline
        & p254 & p236 & p270 & p282 & p271 & p264 & p345 & p294 \\ \hline
        Gender & male & female & male & female & male & female & male & female \\ 
        Accents & English & English & English & English & Scottish & Scottish & American & American \\
        10 utterances in seconds (s) & 37 & 26 & 27 & 33 & 37 & 38 & 31 & 34 \\
        320 utterances in minutes (min) & 15.7 & 13.9 & 16.1 & 19.2 & 18.3 & 19.2 & 16.9 & 18.4 \\ \hline
    \end{tabular}}
    \label{tab:detailinfo}
\vspace{-3mm}
\end{table*}

\subsubsection{Subjective evaluation}
Natural samples and synthetic samples generated by various systems for eight target speakers were evaluated in terms of quality and similarity values. The target models were obtained using either 10 or 320 utterances of the target speakers. 25 native English speakers were asked to answer 26 quality and 26 similarity questions in one session. We collected answers from 229 sessions in the end, with each participant having done about 10 sessions.

\begin{table}[tb]
    \caption{Quality result of individual speaker (5-points scale)}
    \centering
    \scalebox{0.94}{
    \setlength{\tabcolsep}{5pt}
    \begin{tabular}{lrrrrrrrr}
        \hline \hline
        & p254 & p236 & p270 & p282 & p271 & p264 & p345 & p294 \\ \hline
        
        \multicolumn{9}{l}{Reference:} \\
        \hspace{3mm}\texttt{NAT} & \textbf{4.78} & \textbf{4.82} & \textbf{4.31} & \textbf{4.84} & \textbf{4.44} & \textbf{4.75} & \textbf{4.78} & \textbf{4.88} \\
        
        \multicolumn{9}{l}{Adapt with 10 utterances:} \\
        \hspace{3mm}\texttt{MU-A1B} & 2.42 & 3.35 & 2.58 & 3.12 & 2.57 & 2.88 & 2.72 & 2.30 \\
        \hspace{3mm}\texttt{WO-i} & 2.18 & 2.48 & 1.71 & 2.20 & 1.84 & 2.44 & 2.34 & 1.93 \\
        \hspace{3mm}\texttt{BaB}$_\texttt{s}$ & 2.23 & 3.37 & 2.52 & 3.07 & 2.40 & 3.05 & 2.30 & 1.96 \\
        \hspace{3mm}\texttt{BaB}$_\texttt{u}$ & 2.36 & 3.11 & \textbf{2.68} & 3.10 & \textbf{2.95} & 3.32 & 2.86 & 2.30 \\
        \hspace{3mm}\texttt{BaB}$^\texttt{all}_\texttt{s+}$ & 2.70 & \textbf{3.55} & 2.55 & 3.38 & 2.63 & \textbf{3.40} & 2.67 & 2.67 \\
        \hspace{3mm}\texttt{BaB}$^\texttt{all}_\texttt{u}$ & \textbf{2.91} & 3.48 & 2.58 & \textbf{3.40} & 2.90 & 3.21 & \textbf{2.91} & \textbf{2.95} \\
        
        \multicolumn{9}{l}{Adapt with 320 utterances:} \\
        \hspace{3mm}\texttt{MU-A1B} & 2.61 & 3.33 & 2.19 & 3.13 & 2.68 & 3.35 & 2.79 & 2.78 \\ 
        \hspace{3mm}\texttt{WO-i} & 2.02 & 2.44 & 1.79 & 2.21 & 1.69 & 2.19 & 2.16 & 2.16 \\ 
        \hspace{3mm}\texttt{BaB}$_\texttt{s}$ & 2.48 & 3.24 & 2.54 & 3.21 & 2.61 & 3.00 & 3.09 & 2.77 \\
        \hspace{3mm}\texttt{BaB}$_\texttt{u}$  & 2.78 & 3.23 & 2.81 & 3.40 & 2.87 & 3.35 & 2.77 & 3.17 \\ 
        \hspace{3mm}\texttt{BaB}$^\texttt{all}_\texttt{s+}$ & 3.07 & 3.71 & 2.65 & 3.53 & 2.95 & 3.36 & 3.45 & \textbf{3.35} \\ 
        \hspace{3mm}\texttt{BaB}$^\texttt{all}_\texttt{u}$ & \textbf{3.44} & \textbf{3.77} & \textbf{3.32} & \textbf{3.70} & \textbf{3.03} & \textbf{3.62} & \textbf{3.48} & 3.29 \\ \hline
    \end{tabular}}
    \label{tab:detailqua}
\end{table}

The subjective results are showed in Fig. \ref{fig:vctk010} for the 10-utterance case and in Fig. \ref{fig:vctk320} for the 320-utterance case. Comparing \texttt{WO-i} with the other strategies, we can see that systems using the WaveNet vocoder are overwhelmingly better than WORLD especially in the quality measurement.
Both supervised and unsupervised baselines (\texttt{MU-A1B} and \texttt{WO-i}) failed to show any significant improvement when the training data was increased from 10 utterances to 320 utterances as shown in Fig. \ref{fig:vctk010} and \ref{fig:vctk320}. In contrast, the proposed strategies get a big boost when more data are available. This is one of the advantages of our method over the existing approaches.
Among the proposed strategies, $\texttt{BaB}^\texttt{all}$ is still better than $\texttt{BaB}$ similar to the results of previous experiment with the Japanese corpus. The unsupervised systems are slightly better than the supervised in term of quality while it is reversed for similarity measurement. The auxiliary tactic applied to $\texttt{BaB}^\texttt{all}_\texttt{s+}$ does not seem to help it surpass its unsupervised counterpart $\texttt{BaB}^\texttt{all}_\texttt{u}$.
A more sophisticated tactic with a similar nature might yeild further improvements.

Finally, the detailed results for individual speakers are listed in Table \ref{tab:detailqua} and \ref{tab:detailsim}. In terms of quality, the unsupervised system \texttt{BaB}$^\texttt{all}_\texttt{u}$ has consistently superior performance when it is adapted with 320 utterances. However, with 10 utterances, the supervised system \texttt{BaB}$^\texttt{all}_\texttt{s+}$ or the conservative adaptation approach \texttt{BaB}$_\texttt{u}$ has better results for certain speakers. Interestingly, while the unsupervised system \texttt{BaB}$^\texttt{all}_\texttt{u}$ has the best quality, the similarity results are split between \texttt{BaB}$^\texttt{all}_\texttt{u}$ and \texttt{BaB}$^\texttt{all}_\texttt{s+}$. 
One hypothesis is that while \texttt{BaB}$^\texttt{all}_\texttt{u}$ can always secure natural-sounding samples, it fails to establish a correct mapping between the linguistic representation and the actual pronunciation of the target speakers as the adaptation process does not include the linguistic encoders.
As shown in Table \ref{tab:detailsim}, \texttt{BaB}$^\texttt{all}_\texttt{u}$ has the worst similarity scores for p264 and p345 who happen to be speakers with non-English accents. However, any conclusion at present would be speculation; more experiments will be needed before we can make a comprehensive conclusion about performance for speakers with varying characteristics.

\begin{table}[tb]
    \caption{Similarity result of individual speaker (4-points scale)}
    \centering
    \scalebox{0.94}{
    \setlength{\tabcolsep}{5pt}
    \begin{tabular}{lrrrrrrrr}
        \hline \hline
        & p254 & p236 & p270 & p282 & p271 & p264 & p345 & p294 \\ \hline
        
        \multicolumn{9}{l}{Reference:} \\
        \hspace{3mm}\texttt{NAT} & \textbf{3.97} & \textbf{3.86} & \textbf{3.90} & \textbf{3.90} & \textbf{3.85} & \textbf{3.93} & \textbf{3.96} & \textbf{3.98} \\
        \multicolumn{9}{l}{Adapt with 10 utterances:} \\
        \hspace{3mm}\texttt{MU-A1B} & 2.32 & 2.34 & 2.40 & 2.63 & 2.81 & 2.16 & 2.57 & 2.77 \\
        \hspace{3mm}\texttt{WO-i} & 2.04 & 2.67 & 2.42 & 2.33 & 2.59 & 2.50 & 2.28 & 2.10 \\
        \hspace{3mm}\texttt{BaB}$_\texttt{s}$ & 2.49 & 2.62 & 2.58 & 2.93 & 2.83 & \textbf{2.73} & 2.25 & 2.40 \\
        \hspace{3mm}\texttt{BaB}$_\texttt{u}$ & 2.33 & 2.74 & 2.91 & 2.71 & 3.09 & 2.33 & 2.62 & 2.43 \\
        \hspace{3mm}\texttt{BaB}$^\texttt{all}_\texttt{s+}$ & 2.88 & 3.02 & \textbf{3.02} & \textbf{3.02} & \textbf{3.37} & 2.40 & \textbf{2.85} & \textbf{2.95} \\
        \hspace{3mm}\texttt{BaB}$^\texttt{all}_\texttt{u}$ & \textbf{2.98} & \textbf{3.04} & 2.93 & 2.90 & 3.07 & 2.41 & 2.68 & 2.80 \\
        
        \multicolumn{9}{l}{Adapt with 320 utterances:} \\
        \hspace{3mm}\texttt{MU-A1B} & 2.22 & 2.37 & 2.81 & 2.79 & 2.83 & 2.29 & 2.61 & 2.78 \\
        \hspace{3mm}\texttt{WO-i} & 2.19 & 2.75 & 2.38 & 2.32 & 2.83 & 2.51 & 2.26 & 2.28 \\
        \hspace{3mm}\texttt{BaB}$_\texttt{s}$ & 2.81 & 3.15 & 2.95 & 3.14 & 3.12 & \textbf{3.40} & 2.75 & 2.96 \\ 
        \hspace{3mm}\texttt{BaB}$_\texttt{u}$ & 2.62 & 3.11 & 2.88 & 2.91 & 3.05 & 3.03 & 2.63 & \textbf{3.19} \\ 
        \hspace{3mm}\texttt{BaB}$^\texttt{all}_\texttt{s+}$ & 3.37 & \textbf{3.55} & \textbf{3.05} & 3.12 & 3.21 & 3.32 & \textbf{3.66} & 3.11 \\
        \hspace{3mm}\texttt{BaB}$^\texttt{all}_\texttt{u}$ & \textbf{3.45} & 3.34 & 3.00 & \textbf{3.19} & \textbf{3.51} & 2.86 & 2.98 & 3.12 \\ \hline

    \end{tabular}}
    \label{tab:detailsim}

\end{table}

\section{Conclusion}
\label{sec:conclusion}
We systematically reviewed the methodology of speaker adaptation for speech synthesis systems and pointed out the remaining limitations. We then proposed a unified framework for conducting supervised and unsupervised adaptation with backpropagation.
Our method can use different types of speaker components to model the speaker transformation instead of assuming that the speaker characteristics can be represented by a single fixed-length vector.
Further this approach allows us to fine tune the entire acoustic decoder even if the adaptation data do not include transcriptions.
The results of the experiments suggest that by providing a good initial factorized model, fine-tuning the entire acoustic decoder yields the best performance for both supervised and unsupervised adaptation.

Interestingly, the unsupervised adapted model turned out to be significantly better than its supervised counterpart in the subjective test. Our hypothesis is that element-wise metrics like the mean square error might not reflect human perception. A similar conclusion has been suggested in other studies involving speech \cite{saito2018statistical} and image generation \cite{larsen2015autoencoding}. Incorporating a generative adversarial network (GAN) \cite{saito2018statistical} into the architecture is a popular way to address this issue.
The performance of the adapted model still varies between target speakers. Our future work will focus on improving the performance and robustness of the proposed method.

\section*{Acknowledgment}

This work was partially supported by a JST CREST Grant (JPMJCR18A6, VoicePersonae project), Japan, and MEXT KAKENHI Grants (16H06302, 17H04687, 18H04120, 18H04112, 18KT0051), Japan. 
We are grateful to Dr. Erica Cooper for helpful comments.

\ifCLASSOPTIONcaptionsoff
  \newpage
\fi

\bibliographystyle{IEEEtran}
\bibliography{main}

\newpage

\begin{IEEEbiography}{Hieu-Thi Luong}
received the B.E and M.E degrees in computer science from Vietnam National University, Ho Chi Minh city, University of Science, Vietnam in 2014 and 2016 respectively. Since 2017, he is awarded the Japanese Government (Monbukagakusho: MEXT) Scholarship to pursue a PhD degree on statistical speech synthesis and machine learning at National Institute of Informatics, Tokyo, Japan.
\end{IEEEbiography}

\begin{IEEEbiography}{Junichi Yamagishi}
received the Ph.D. degree from Tokyo Institute of Technology in 2006 for a thesis that pioneered speaker-adaptive speech synthesis. He is currently a Professor with the National Institute of Informatics, Tokyo, Japan, and also a Senior Research Fellow with the Centre for Speech Technology Research, University of Edinburgh, Edinburgh, U.K. Since 2006, he has authored and co-authored more than 250 refereed papers in international journals and conferences. 

He was the recipient of the Tejima Prize as the best Ph.D. thesis of Tokyo Institute of Technology in 2007. He was awarded the Itakura Prize from the Acoustic Society of Japan in 2010, the Kiyasu Special Industrial Achievement Award from the Information Processing Society of Japan in 2013, and the Young Scientists' Prize from the Minister of Education, Science and Technology in 2014, the JSPS Prize from Japan Society for the Promotion of Science in 2016, and Docomo mobile science award from Mobile communication fund in 2018. 

He was one of organizers for special sessions on ``Spoofing and Countermeasures for Automatic Speaker Verification" at Interspeech 2013, ``ASVspoof evaluation" at Interspeech 2015, ``Voice conversion challenge 2016" at Interspeech 2016, and ``2nd ASVspoof evaluation" at Interspeech 2017, ``Voice conversion challenge 2018” at Speaker Odyssey 2018. He is currently an organizing committee for ASVspoof 2019, an organizing committee for ISCA the 10th ISCA Speech Synthesis Workshop 2019, a technical program committee for IEEE ASRU 2019, and an award committee for ISCA Speaker Odyssey 2020.

He was a member of the Speech and Language Technical Committee and a Lead Guest Editor for the IEEE JOURNAL OF SELECTED TOPICS IN SIGNAL PROCESSING special issue on Spoofing and Countermeasures for Automatic Speaker Verification. He is currently a Senior Area Editor of the IEEE/ACM TRANSACTIONS ON AUDIO, SPEECH, AND LANGUAGE PROCESSING and a chairperson of ISCA SynSIG.
\end{IEEEbiography}

\end{document}